\documentclass{aa}

\usepackage{graphicx}

\usepackage{newtxtext}
\usepackage{hyperref}
\hypersetup{breaklinks=true}
\hypersetup{colorlinks,                                 
        linkcolor={red!50!black},
        citecolor={blue!50!black},
        urlcolor={blue!80!black}}
\usepackage{lipsum}                     
\usepackage{xargs}                      
\usepackage[pdftex,dvipsnames]{xcolor}  

\usepackage[colorinlistoftodos,prependcaption,textsize=tiny]{todonotes}
\usepackage{fdsymbol}
\usepackage{booktabs}
\usepackage{placeins}
\usepackage{siunitx}

\usepackage{caption} 
\usepackage[nameinlink]{cleveref}
\usepackage{pdfpages}
\usepackage{orcidlink}
\usepackage{amsmath}

\newcommand{\Na}{Na\,{\textsc{i}}}
\newcommand{\Ha}{H\,{\textsc{$\alpha$}}}

\newcommand{\Naupperlimvaluewounit}{$3.34\%$}
\newcommand{\Haupperlimvaluewounit}{$3.97\%$}
\newcommand{\Naupperlimplanrad}{$3\sigma_{lim}=1.8R_p$}
\newcommand{\Haupperlimplanrad}{$3\sigma_{lim}=1.9R_p$}
\newcommand{\Naupperlimplanradwosigma}{$1.8R_p$}
\newcommand{\Haupperlimplanradwosigma}{$1.9R_p$}

\newcommand{\obliquity}{$\lambda = -5.2 \pm 3.4 ^\circ$}
\newcommand{\stelprojvel}{$v_{\rm eq} \sin{i_\star} = 2.58 \pm 0.12$ kms$^{-1}$}

\begin{document}                        

   \title{Hot Exoplanet Atmospheres Resolved with Transit Spectroscopy (HEARTS)}

   \subtitle{VIII. Nondetection of sodium in the atmosphere of the aligned planet KELT-10b\thanks{Based on observations made at the ESO 3.6m telescope (La Silla, Chile) under ESO programme 097.C-1025}}

   \author{
   M. Steiner\inst{1}\orcidlink{0000-0003-3036-3585}\thanks{\href{mailto:Michal.Steiner@unige.ch}{Michal.Steiner@unige.ch}} \and 
   O. Attia\inst{1}\orcidlink{0000-0002-7971-7439} \and 
   D. Ehrenreich\inst{1} \and 
   M. Lendl\inst{1} \and 
   V. Bourrier\inst{1} \and 
   C. Lovis\inst{1} \and 
   J. V. Seidel\inst{2}\orcidlink{0000-0002-7990-9596} \and 
   S. G. Sousa\inst{3} \and 
   D. Mounzer\inst{1} \and 
   N. Astudillo-Defru\inst{4} \and 
   X. Bonfils\inst{5} \and 
   V. Bonvin\inst{6} \and 
   W. Dethier\inst{5} \and 
   K. Heng\inst{7} \and 
   B. Lavie\inst{1} \and 
   C. Melo\inst{8} \and 
   G. Ottoni\inst{1}\orcidlink{0000-0001-9305-9631} \and 
   F. Pepe\inst{1} \and 
   D. S\'egransan\inst{1} \and 
   A. Wyttenbach\inst{1}\orcidlink{0000-0001-9003-7699} 
}

   \institute{
    Observatoire de l’Universit\'e de Gen\`eve, Chemin Pegasi 51, 1290 Versoix, Switzerland
         \and
    European Southern Observatory, Alonso de Córdova 3107, Vitacura, Región Metropolitana, Chile
         \and
    Instituto de Astrofísica e Ciências do Espaço, Universidade do Porto, CAUP, Rua das Estrelas, 4150-762 Porto, Portugal 
         \and
    Departamento de Matemática y Física Aplicadas, Universidad
Católica de la Santísima Concepción, Alonso de Rivera 2850,
Concepción, Chile 
        \and
    Université Grenoble Alpes, CNRS, IPAG, 38000 Grenoble, France 
         \and
    NetGuardians SA  
        \and
        Ludwig Maximilian University, University Observatory Munich, Scheinerstrasse 1, Munich 81679, Germany
        \and
    Portuguese Space Agency, Estrada das Laranjeiras, n.º 205, RC,
1649-018, Lisboa, Portugal 
    }
   \date{Received ...; accepted ...}

 
  \abstract
   {
   The HEARTS survey aims to probe the upper layers of the atmosphere by detecting resolved sodium doublet lines, a tracer of the temperature gradient, and atmospheric winds. KELT-10b, one of the targets of HEARTS, is a hot-inflated Jupiter with 1.4 $R_\mathrm{Jup}$ and 0.7 $M_\mathrm{Jup}$. Recently, there was a report of sodium absorption in the atmosphere of KELT-10b (0.66\% ± 0.09\% (D2) and 0.43\% ± 0.09\% (D1); VLT/UVES data from single transit).
   }
   {We searched for potential atmospheric species in KELT-10b, focusing on sodium doublet lines (\Na; 589 nm) and the Balmer alpha line (\Ha; 656 nm) in the transmission spectrum. Furthermore, we measured the planet-orbital alignment with the spin of its host star.
   }
   {We used the Rossiter-McLaughlin Revolutions technique to analyze the local stellar lines occulted by the planet during its transit. We used the standard transmission spectroscopy method to probe the planetary atmosphere, including the correction for telluric lines and the Rossiter-McLaughlin effect on the spectra. We analyzed two new light curves jointly with the public photometry observations. }
  {We do not detect signals in the \Na\,and \Ha\,lines within the uncertainty of our measurements. We derive the $3\sigma$ upper limit of excess absorption due to the planetary atmosphere corresponding to equivalent height $R_p$ to \Naupperlimplanradwosigma\,(\Na) and \Haupperlimplanradwosigma\,(\Ha). The analysis of the Rossiter-McLaughlin effect yields the sky-projected spin-orbit angle of the system \obliquity and the stellar projected equatorial velocity \stelprojvel. Photometry results are compatible within $1\sigma$ with previous studies.}
   {We found no evidence of \Na\,and \Ha, within the precision of our data, in the atmosphere of KELT-10b. Our detection limits allow us to rule out the presence of neutral sodium or excited hydrogen in an escaping extended atmosphere around KELT-10b. We cannot confirm the previous detection of \Na\,at lower altitudes with VLT/UVES. We note, however, that the Rossiter-McLaughlin effect impacts the transmission spectrum on a smaller scale than the previous detection with UVES. Analysis of the planet-occulted stellar lines shows the sky-projected alignment of the system, which is likely truly aligned due to tidal interactions of the planet with its cool (Teff < 6250 K) host star.
  }


   \maketitle
%
\section{Introduction}
\label{sec_1}
Shortly after the discovery of the first exoplanet orbiting a solar-type star, 51 Peg b \citep{mayor1995}, we were able to observe the atmosphere around a similar exoplanet (HD 209458b) for the first time by detecting sodium through transmission spectroscopy using a space telescope, specifically the Hubble Space Telescope (HST) \citep{charbonneau2002}. This feature was later observed using data from a ground-based facility by the Subaru telescope/High Dispersion Spectrograph (HDS) \citep{snellen2008}. Since then, the field of exoplanet atmospheres has moved from the simple detection of chemical species to the characterization of atmospheric dynamics. This includes observations of wind patterns \cite[e.g.,][]{seidel2020b,seidel2021} or the asymmetry between the morning and evening terminators detected by \cite{ehrenreich2020} and \cite{kesseli2021}. Furthermore, we can observe not only the strong resonant lines such as that of the sodium doublet \citep{seidel2019,langeveld2022} but also target the cross-correlation function of the weak lines using spectra from high-resolution spectrographs such as High Accuracy Radial velocity Planet Searcher (HARPS) \citep{mayor2003}, Echelle SPectrograph for Rocky Exoplanets and Stable Spectroscopic Observations (ESPRESSO) \citep{pepe2021}, Ultraviolet and Visual Echelle Spectrograph (UVES) \citep{dekker2000}, or HDS \citep{noguchi2002,sato2002}. Using the latter method allows for the creation of a "spectroscopic atlas," as shown, for example, in \cite{hoeijmakers2019,hoeijmakers2020}.

While the field is rapidly gaining momentum through the use of larger telescopes, we have also started to realize the impact of stellar and telluric effects on the results of transmission spectroscopy. The first sodium detection \citep{charbonneau2002} has been recently disputed by \cite{casasayas-barris2021}; the reported signal seems to come from the Rossiter-McLaughlin (RM) effect \cite[see also][]{sing2008,casasayas-barris2020}. RM is an effect where we observe a radial velocity (RV) anomaly from the expected Keplerian value caused by the stellar rotation and transit of the planet. Due to the stellar rotation, the planetary disk occults regions of the stellar surface with different velocities and spectra properties. Since the RV is derived from the disk-integrated spectrum of the star, which is deformed by the occulted regions, we observe the velocity anomaly. In the case of ground-based observations, the telluric lines can also cause false positives and negatives \citep[e.g.,][]{seidel2020,seidel2020a,langeveld2021a,orell-miquel2022,spake2022}, and a careful correction of the Earth's atmosphere is necessary.

The scientific exploration of exoplanetary atmospheres, namely the detection and characterization of chemical composition, is becoming more and more mainstream \citep[e.g.,][]{borsa2021,chen2020,deibert2021,langeveld2022}, such that we can start to perform population studies. The population sample size is mainly limited by the brightness of the host star and atmospheric scale height. There are also many potential exoplanets for planetary characterization not observed by any sufficient observatory yet.

Recently, several teams have attempted to summarize and characterize the detected species in exoplanetary atmospheres, including \citet[][Table 1]{guillot2022} and \citet[][Figure 7]{stangret2021}, for example. The former summarize chemical species identified in exoplanetary atmospheres, highlighting the size of the studied sample. The latter show species detected based on planets with $T_{eq} >1000K$, mainly focusing on (ultra)hot Jupiters. Atomic species, such as sodium or hydrogen, were observed only in several tens of exoplanetary atmospheres, sampling mainly hot Jupiters and warm Neptunes.

\begin{figure}
    \centering
    \includegraphics[width=\hsize]{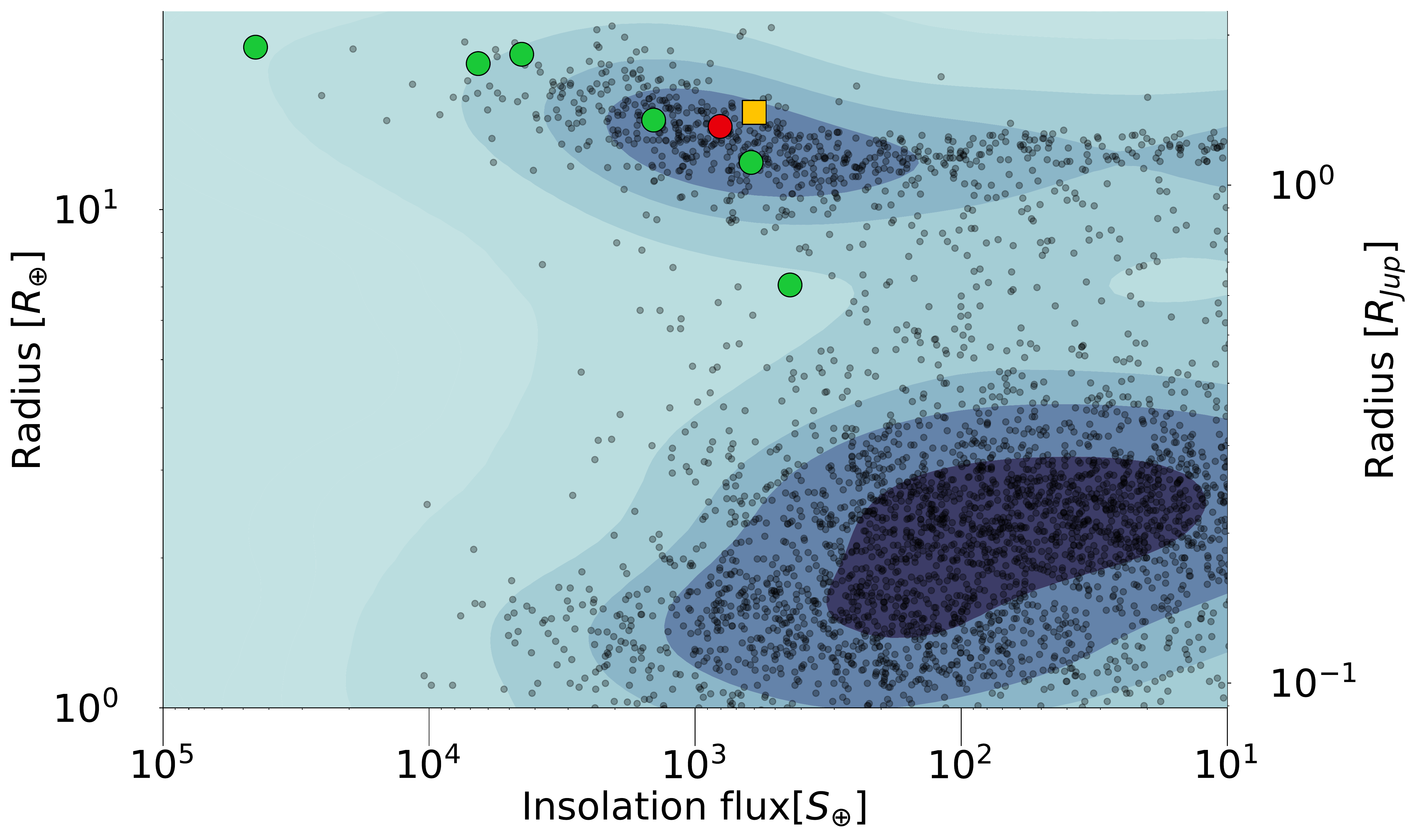}
    \caption{Radius-insolation flux figure of known exoplanets. The color-coding used is as follows: yellow, KELT-10b; green, HEARTS and SPADES surveys' sodium detections (SPADES is a twin survey using a similar setup as HARPS-N at the 3.58m telescope at La Palma); and red, WASP-127, the only nondetection from HEARTS. WASP-127 b has not been detected in HARPS data \citep{seidel2020a}. However, by using VLT/ESPRESSO, \cite{allart2020} established a tenuous signal below the detection limit of \cite{seidel2020a}.
    }
    \label{Fig_radius_insolation}
\end{figure}

We could perform a population study to understand why some chemical species are (non)detectable in the exoplanetary atmosphere. The goal of the Hot Exoplanet Atmospheres Resolved with Transit Spectroscopy (HEARTS) survey is to characterize a sample of hot Jupiters with the high-resolution spectrograph HARPS \citep{mayor2003} \autoref{Fig_radius_insolation}, \citep{wyttenbach2017,seidel2019,bourrier2020,hoeijmakers2020,seidel2020,seidel2020a,mounzer2022}, and to serve as a pathfinder for VLT/ESPRESSO \citep{pepe2021}. Data from HEARTS and other surveys are now used for such population studies, such as \cite{langeveld2022}. For the purpose of population study, a set of planets with very similar properties, such as planetary radius and stellar type, is interesting for atmospheric study. However, the sample of currently known exoplanets makes it difficult for such studies to be successful, as not many planets are similar enough and are orbiting stars bright enough to perform such a task.

The target of this paper is KELT-10 b \citep{kuhn2016}. This hot Jupiter has many properties similar to the benchmark planet system HD 209458 b, providing a valuable opportunity to compare these two planets. It notably features a sodium detection obtained by \cite{mccloat2021} with VLT/UVES. We obtained two transits with HARPS as part of HEARTS, which is further discussed in \Cref{sec_2}, with simultaneous photometry observations discussed and analyzed in \Cref{sec_3}. The RM effect analysis is discussed in \Cref{sec_RM}, followed by transmission spectroscopy in \cref{sec_4}, searching for sodium (\Na) and hydrogen (\Ha). Finally, we discuss our results with concluding remarks in \Cref{sec_5}.

\section{Observations}
\label{sec_2}
We observed two transits of KELT-10b on 26 June 2016 and 21 July 2016 from La Silla (Chile). We obtained 96 spectra\footnote{Data available from the \href{http://archive.eso.org/cms.html}{ESO archive}.} with HARPS \citep{mayor2003} at the 3.6 m telescope and two photometric series with EulerCAM (ECAM) at the 1.2 m Euler telescope\footnote{\url{https://www.eso.org/public/teles-instr/lasilla/swiss/}} simultaneously. We refer the reader to \cite{lendl2012} for more information about ECAM. We observed two new light curves that were analyzed jointly with already reduced light curves introduced by \cite{kuhn2016} to refine the ephemeris system.

\begin{table*}
\centering
\begin{tabular}{lrllllll}
\hline
Night &  \# & \# of spectra &         Airmass &          Seeing &            SN56 &            SN67  & Exposure time \\
\hline
2016-06-26 &  1 &   40 (12) &  1.05-1.35-2.17 &  1.12-1.55-2.20 &  17.2-34.5-46.1 &  17.8-34.0-45.0 & 900 s\\
2016-07-21 &  2 &   56 (21) &  1.05-1.31-2.24 &  0.56-0.76-1.08 &  $\mspace{8mu}$8.3-15.8-27.7 &   $\mspace{8mu}$8.0-14.9-26.4 & 600 s \\
\hline
\hline
\end{tabular}
\caption{Observational log: The columns correspond (from left to right) to: Date of observation; Reference number for the given night; Number of spectra during night and number of in-transit spectra in parentheses; Airmass; Seeing and Signal-to-Noise ratio (SNR) for a given order; Exposure time. Note that for the Airmass, Seeing, and SNR columns, the format is min-mean-max value during the night. We refer the reader to \autoref{Fig_night_1} and \autoref{Fig_night_2} for detailed plot.}

\label{tab_obs_log}
\end{table*}

The observational conditions (\autoref{tab_obs_log}) were not ideal either night, with high seeing for the first night and partial cloud cover for the second night. Due to the passage of clouds and exposure time (600s) used on the second night, the observed spectra of this night suffer from a low signal-to-noise ratio (S/N, mean $\sim$15). We ultimately rejected this night from our analysis due to the data quality.

\begin{figure}
    \centering
    \includegraphics[width=\hsize]{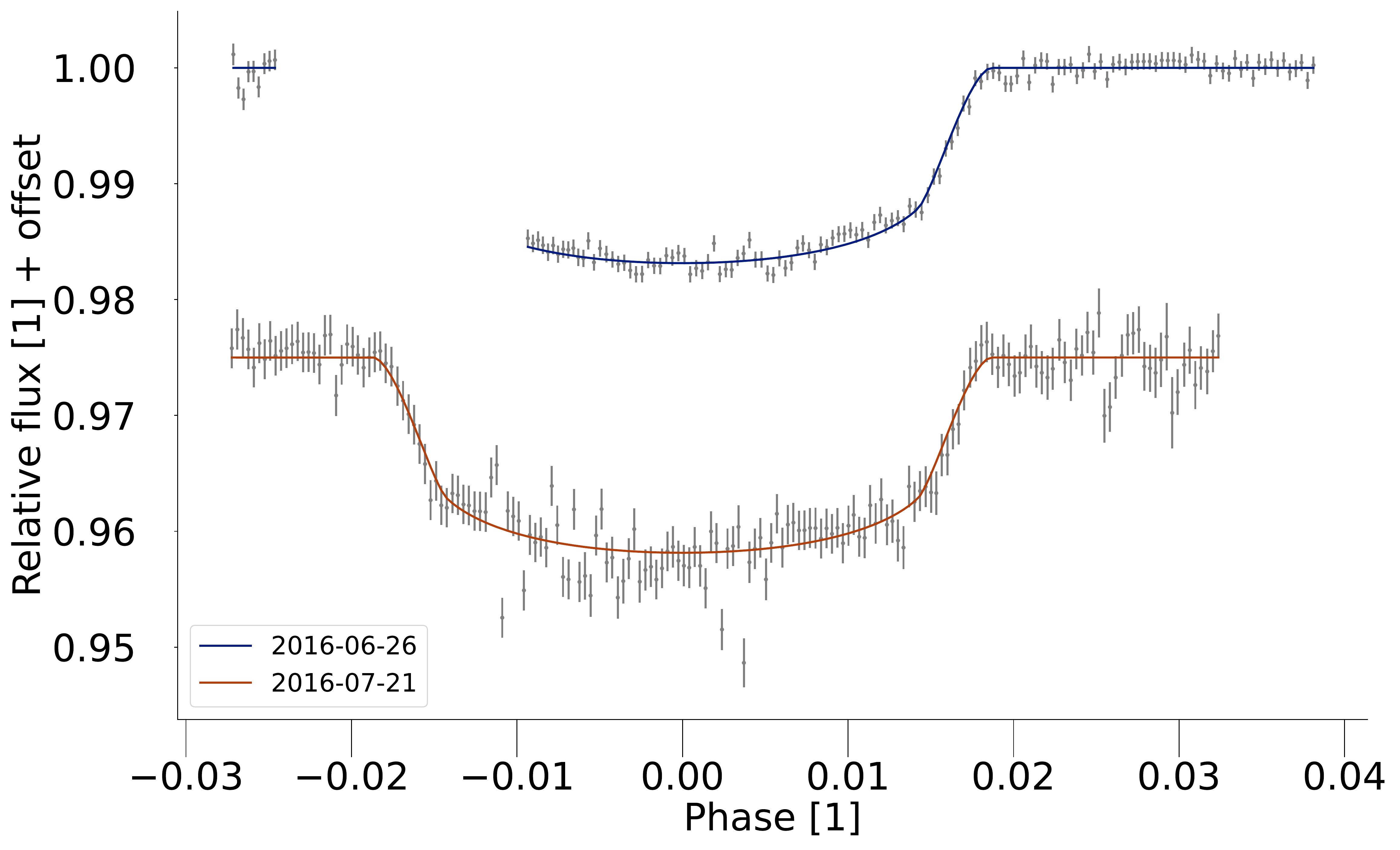}
    \caption{Light curves observed complementary to the HARPS dataset using the ECAM at the 1.2 meter Euler Swiss telescope.}
    \label{Fig_light_curves}
\end{figure}

The ECAM light curves are shown in \autoref{Fig_light_curves}. The entire transit duration is covered, with a short gap due to a telescope software issue during the beginning of transit on night 1. During the second night, a passage of clouds impacted the observations, leading to significant scatter during the transit.

\begin{figure}
    \centering
    \includegraphics[width=\hsize]{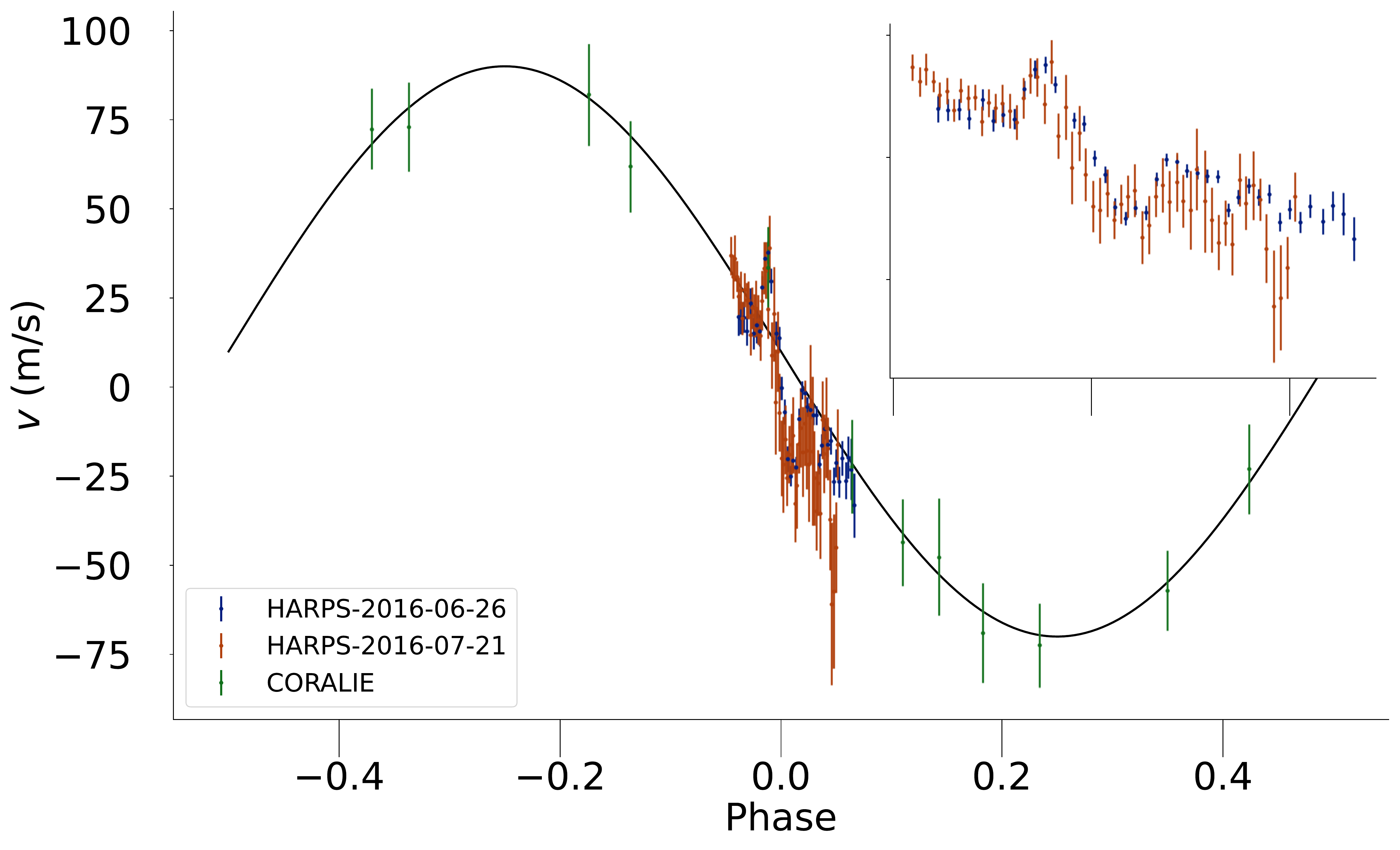}
    \caption{RV observations from CORALIE (green) and HARPS (blue and orange for nights 1 and 2, respectively, analyzed dataset). The black line is a Keplerian model as reported in \cite{kuhn2016}. The inset plot in the upper right corner zooms in on the transit in detail. A strong and roughly symmetrical velocity anomaly due to the RM effect is observed during transit.
    }
    \label{Fig_RV_observation}
\end{figure}

Finally, the CORALIE \citep{queloz2000} RV follow-up from \cite{kuhn2016} has also been used. CORALIE is a spectrograph located at the same telescope as ECAM, with RV measurement precision up to $3$ ms$^{-1}$. CORALIE has been monitoring KELT-10 for potential companions and measuring the planetary mass of KELT-10b. We show all RV observations for both CORALIE and HARPS in \autoref{Fig_RV_observation}.

\section{Photometry analysis}
\label{sec_3}
While the overall precision of ECAM light curves is better than those used in \cite{kuhn2016}, using only the two transits without the \cite{kuhn2016} set of data provides compatible yet less precise system parameters. For that reason, we opted to analyze the entire collection of publicly available follow-up photometry (except for the discovery light curve from the KELT-South telescope), using the \texttt{CONAN} code \citep{lendl2017a}.

The first step in \texttt{CONAN} is to find a proper baseline model for the light curve. We searched for different trends, including airmass, sky, $x$ and $y$ coordinate shift, Full Width Half Maximum (FWHM), and sinusoidal pattern. Both ECAM light curves were detrended separately, and the appropriate model was selected using the difference in Bayesian Information Criterion ($\Delta$BIC) value and checking by eye (essential to avoid over-fitting). The 26 June 2016 light curve has been detrended for time and FWHM trends, while the 21 July 2016 light curve has been detrended for time and sky trends.

For the set of limb-darkening coefficients, we used the publicly available code LDCU\footnote{Available at: \url{https://github.com/delinea/LDCU}} \citep{deline2022}. The stellar parameters necessary for the limb-darkening computation come from the UVES analysis by \cite{sousa2018}.

Once we established a proper baseline, we ran \texttt{CONAN} on the entire set, using the  differential evolution Markov chain Monte Carlo (DEMCMC) method described in \cite{lendl2017a}. We ran the code multiple times, and it varied whether or not we fixed an initial parameter between the different usable jump parameters. In the end, we fixed semi-amplitude K to the value from \cite{kuhn2016}. We found better results with eccentricity set to 0, adopting these values. All resulting parameters are shown in \autoref{Tab_conan_output}. Finally, we searched for a wavelength dependence of the transit depth in different filters but measured consistent broadband $R_p$. 

\begin{table*}
\bgroup
\resizebox{\textwidth}{!}{%
\def\arraystretch{1.5}

\begin{tabular}{llllc}

\hline
Jump parameters   & Parameter & \cite{kuhn2016}                 & CONAN analysis                         & $\sigma$-- difference    \\ \hline
$T_0$ [$BJD_\mathrm{TBD}$]          & Mid transit time & $2457175.0439^{+0.0053}_{-0.0054}$ & $2457175.04197_{-0.00029}^{+0.00028}$  & 0.36 \\ 
$\frac{R_p}{R_s}$ [1] & Planet to star radius ratio    & $0.1190^{+0.0014}_{-0.0012}$       & $0.11978_{-0.0011}^{+0.00076}$  & 0.44 \\ 
i [deg]     & Inclination & $88.61^{+0.86}_{-0.74}$             & $87.82_{-0.45}^{+0.54}$            & 0.86 \\ 
$\frac{a}{R_s}$ [1]           & System scale & $9.34^{+0.21}_{-0.32}$              & $9.03_{-0.20}^{+0.27}$             &  0.74\\ 
P [d]       &  Period & $4.166285_{-0.000057}^{+0.000057}$ & $4.1662449_{-0.0000032}^{+0.0000032}$ & 0.70\\ 
\hline
Derived parameters& Parameter & \cite{kuhn2016}                & CONAN analysis                   &  $\sigma$-- difference \\ \hline
$R_p$ [$R_\mathrm{Jup}$]  & Planetary radius & $1.399^{+0.069}_{-0.049}$          & $1.409_{-0.051}^{+0.046}$    & 0.12\\ 
$M_p$ [$M_\mathrm{Jup}$]  & Planetary mass  & $0.679^{+0.039}_{-0.038}$          & $0.680_{-0.024}^{+0.024}$    & 0.02\\ 
a [au]            & Semi-major axis & $0.05250^{+0.000 86}_{-0.00097}$   & $0.0508_{-0.0019}^{+0.0024}$ & 0.66\\ 
b [1]            & Impact parameter & $0.23^{+0.11}_{-0.14}$             & $0.344_{-0.075}^{+0.065}$    & 0.86\\ 
$T_{14}$ [day] & Transit duration  & $0.1560^{+0.0016}_{-0.0012}$       & $0.1570_{-0.0013}^{+0.0010}$ & 0.49\\ 
$dF$ or $\frac{R_p^2}{R_s^2}$ [1]& Transit depth & $0.01416^{+0.00034}_{-0.00029}$    & $0.01435_{-0.00026}^{+0.00018}$ & 0.44\\ 
\hline
\hline
\end{tabular}%
}
\egroup
\caption{Comparison of the results from the CONAN analysis and the parameters from \cite{kuhn2016}. The top part of the table includes the parameters that have been used as jump parameters. The bottom part provides additional parameters we can derive from the jump parameters. From left to right, the columns correspond to the following: parameter (and unit); \cite{kuhn2016} parameter set; CONAN analysis (this paper); and $\sigma$ difference between the parameters.}
\label{Tab_conan_output}
\end{table*}

The resulting ephemeris is compatible with \cite{kuhn2016} within $1 \sigma$, which can be attributed to different analysis methods. The precision of $T_0$ and $P$ values has been improved significantly (by a factor of $\sim 18$, new precision of $T_0$ is $^{+0.00028}_{-0.00029}$ and precision of $P$ is $^{+0.0000032}_{-0.0000032}$) due to the more extended baseline of observations.

\section{Rossiter--McLaughlin effect analysis} 
\label{sec_RM}

\begin{figure}
    \centering
    \includegraphics[width=\hsize]{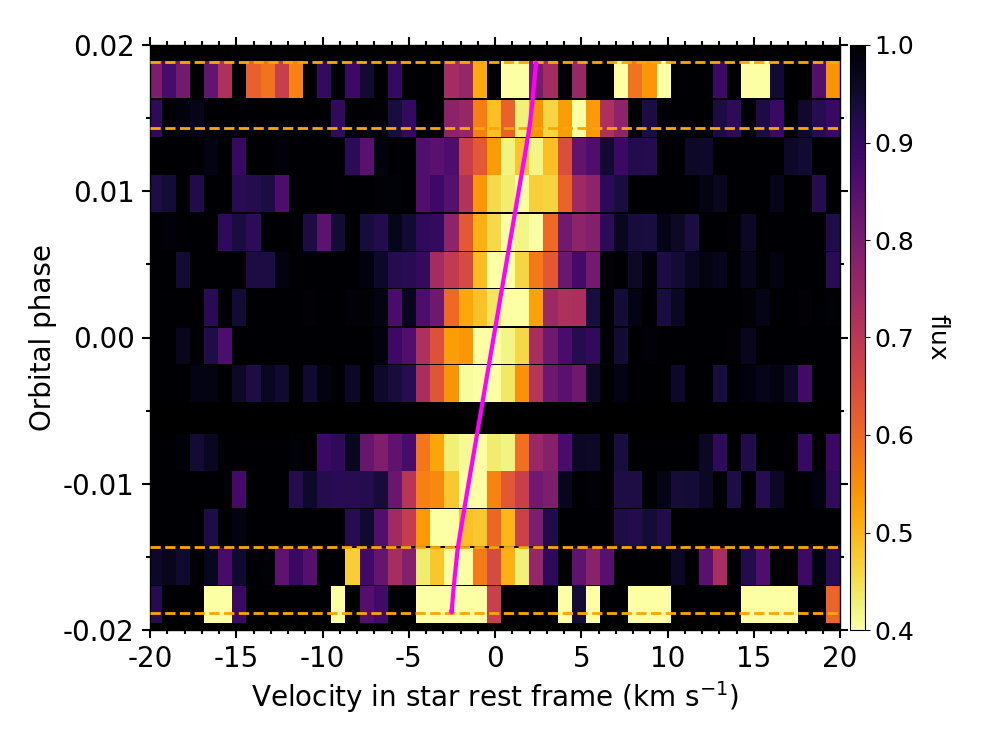}
    \caption{Map of the CCF$_{\rm intr}$ during the transit of KELT-10 b. Transit contacts are shown as orange dashed lines. Values are colored as a function of their normalized flux, and plotted as a function of the RV in the star rest frame (in abscissa) and orbital phase (in ordinate). The core of the stellar line from the planet-occulted regions is clearly visible as a bright streak. The magenta solid line shows the stellar surface RV model from the RMR best fit. The small gap in the first part of the transit (at phase $\sim -0.005$) is due to a missing exposure.
    }
    \label{fig:RM_2D}
\end{figure}

\subsection{Data reduction}
We exploit one dataset obtained with HARPS during the transit of KELT-10 b on 26 June 2016. Forty exposures were acquired (with exposure times of 900 s), among which 14 are in transit, seven are before the transit, and 19 are after the transit. KELT-10 was observed on fiber A, while fiber B was on the sky. The spectra were extracted from the detector images, corrected for, and calibrated using version 3.5 of the Data Reduction Software (DRS) pipeline\footnote{\url{http://www.eso.org/sci/facilities/lasilla/instruments/harps/doc/DRS.pdf}}. One of the DRS corrections concerns the color effect caused by the variability of extinction induced by Earth's atmosphere \citep[e.g.,][]{bourrier2014,wehbe2020}. The flux balance of the KELT-10 spectra was reset to a G2 stellar spectrum template, before they were passed through weighted cross-correlation \citep{baranne1996,pepe2002} with a G2 numerical mask to compute cross-correlation functions (CCFs) with a step of 0.5 kms$^{-1}$. This mask is part of a new set that was built with weights more representative of the photonic error on the line positions, as described in \citet{bourrier2021}. 

As stated in Sect.~\ref{sec_2}, another transit was recorded with HARPS on 21 July 2016. This visit was discarded from the analysis because of the poor S/N. In fact, including this transit introduces spurious line variations that bias the results and for which we could not correct. We refer to Appendix \ref{app:RM} for more details.

Throughout the RM analysis, the posterior probability distributions (PDFs) of free parameters describing models fitted to the data are sampled using \texttt{emcee} \citep{foreman2013}. The number of walkers and the burn-in phase are adjusted based on the degrees of freedom of the considered problem and the convergence of the chains. Best-fit values for these parameters are set to the median of their PDFs, and their 1$\sigma$ uncertainty ranges are defined using highest density intervals (HDIs).

\begin{figure}[h!]
    \centering
    \includegraphics[width=\hsize]{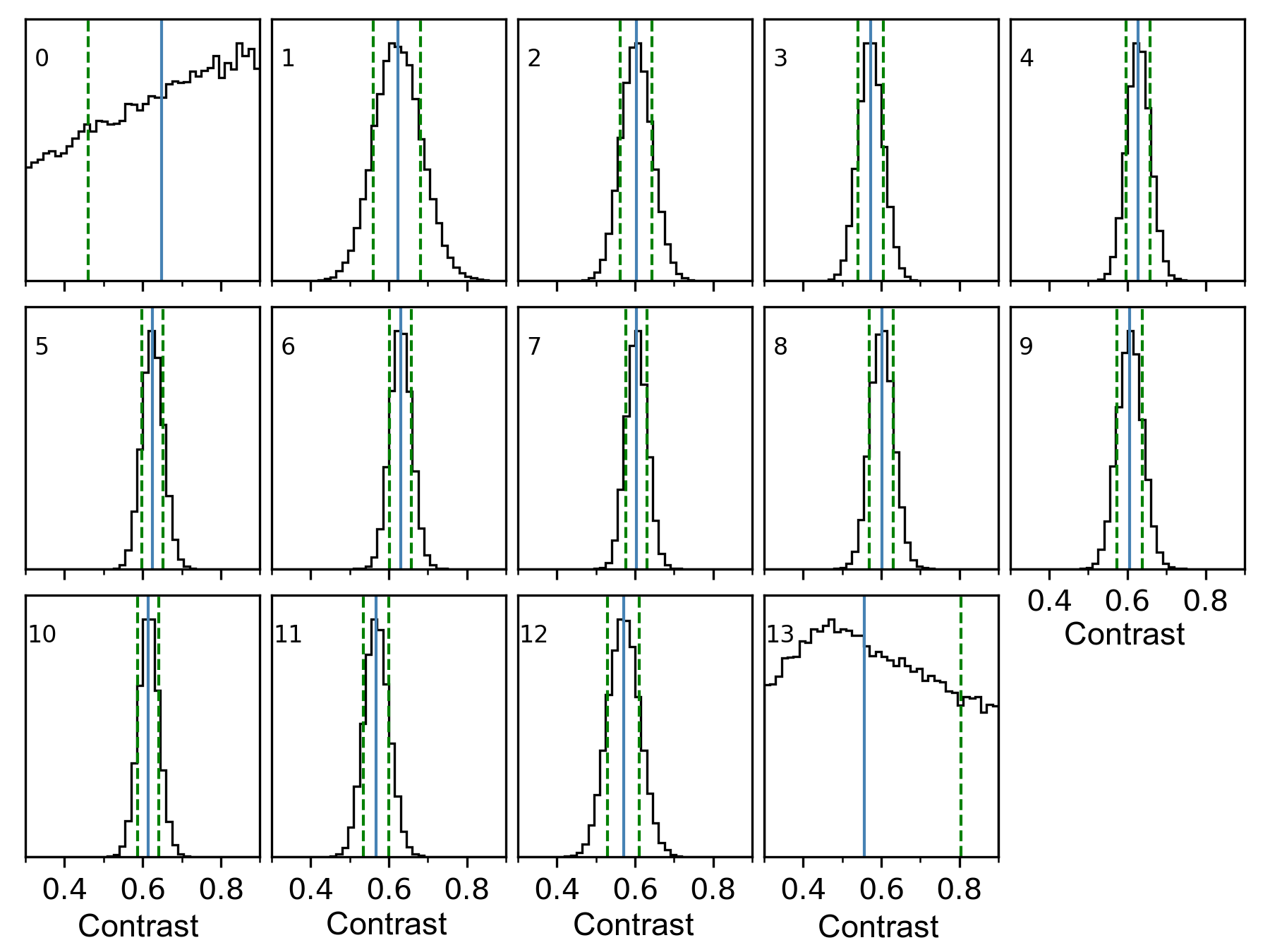}
    \includegraphics[width=\hsize]{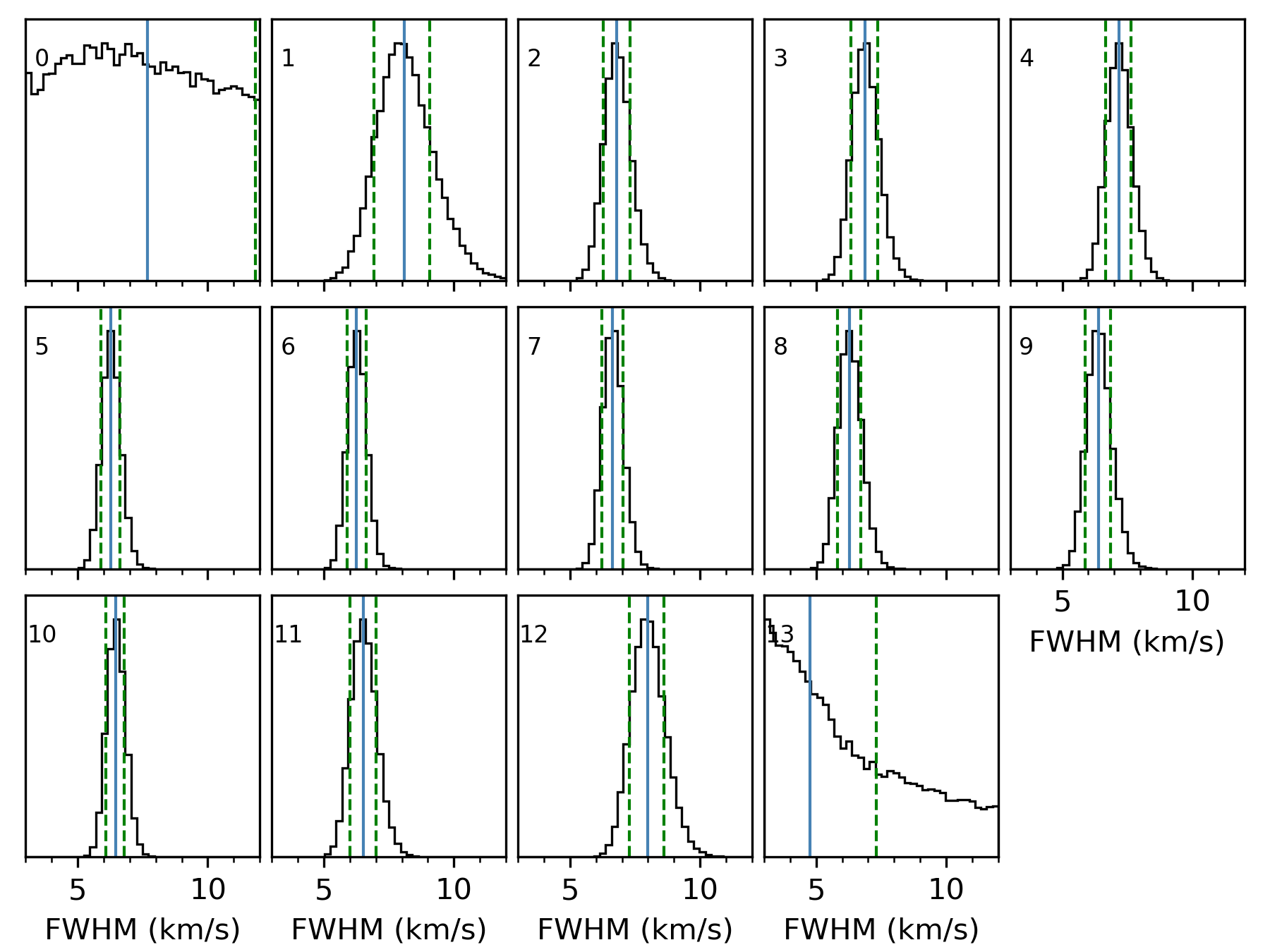}
    \includegraphics[width=\hsize]{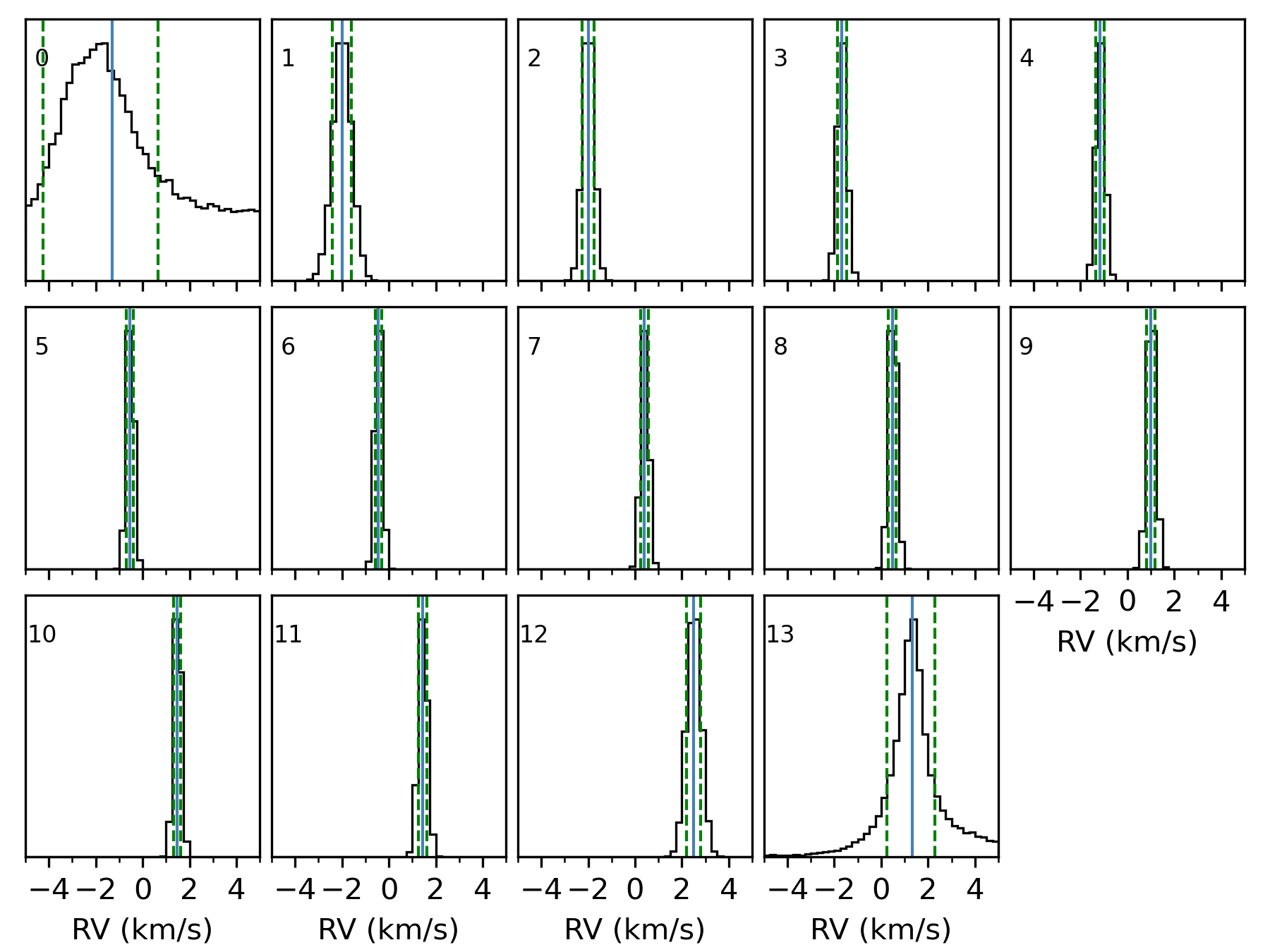}
    \caption{PDFs of the contrast (upper panels), FWHM (middle panels), and RVs (lower panels) of the Gaussian line model fitted to the individual CCF$_{\rm intr}$. Blue lines indicate the PDF median values with green dashed lines showing the 1$\sigma$ highest density interval (HDIs). In-transit exposure indices are shown in each subplot.
    }
    \label{fig:RM_PDF_indiv}
\end{figure}

\subsection{Extraction and analysis of individual exposures}
\label{sec_RM_indiv}

We employed a novel technique, the Rossiter--McLaughlin Revolutions \citep[RMR,][]{bourrier2021} technique, to analyze the data. This method fits the spectral line profiles from all planet-occulted regions at once with a joint model, which boosts the S/N by the number of in-transit exposures at first order, allowing access to fainter RM signals. The RMR is optimal to fully exploit a RM dataset, to assess biases due to line shape variations, and to measure obliquities from all types of systems in a more comprehensive way than traditional techniques. Indeed, a classical RM analysis only relies on the anomalous RVs of the disk-integreated lines, which condense the information contained in CCF profiles into a single measurement and limit our ability to detect the occultation of the stellar surface by the planet. This process can bias our interpretation of the RV anomaly if the occulted stellar line profile is not well modeled or varies along the transit chord. The RMR technique and its holistic analysis of the planet-occulted, rather than the disk-integrated, starlight, has been successfully used to detect the RM signal of the super-Earth HD 3167 b, the smallest exoplanet with a confirmed RM signal \citep{bourrier2021}, and to refine the orbital architecture of the warm Neptune GJ 436 b \citep{bourrier2022}.

The disk-integrated CCF$_{\rm DI}$ of each individual exposure as produced by DRS, corresponding to the light coming from the entire star, are aligned by correcting their velocity table for the Keplerian motion of the star. A master-out CCF, representative of the unocculted star, was built by coadding CCF$_{\rm DI}$ outside of the transit. The master-out CCF is fitted with a Gaussian model, and all CCF$_{\rm DI}$ are shifted to the star rest frame using the derived centroid. Aligned CCF$_{\rm DI}$ are then scaled to a common flux level outside of the transit and to the flux expected from the planetary disk absorption during transit using a light curve computed with the \texttt{batman} package \citep{kreidberg2015}. Cross-correlation functions from the planet-occulted regions are retrieved by subtracting the scaled CCF$_{\rm DI}$ from the master-out CCF and are then reset to a common flux level to yield intrinsic CCF$_{\rm intr}$ that allow for a more direct comparison of the local stellar lines (Fig.~\ref{fig:RM_2D}).

Following the RMR approach, we first analyzed individual exposures to evaluate the quality of each CCF$_{\rm intr}$ and assess the possibility for line variations along the transit chord. The various CCF$_{\rm intr}$ were fitted using a Gaussian model with uniform priors, which  are broad enough to constrain the following parameters: 
$\mathcal{U} (-15,15)$ kms\(^{-1}\) on the RV centroid (i.e., about three times the maximum typical stellar surface RV for G-type stars); \( \mathcal{U} \)$(0,20)$ kms$^{-1}$ on the FWHM (i.e., about three times the width of the master-out CCF, which is assumed to be similar to that of the CCF$_{\rm intr}$ given the slow stellar rotation); and \( \mathcal{U} \)$(0,1)$ on the contrast (i.e., noninformative). A set of 100 walkers were run for 2000 steps, with a burn-in phase of 500 steps. The model was independently fitted to each CCF over [-150, 150] kms$^{-1}$ in the star rest frame. The stellar line is detected in all CCF$_{\rm intr}$ (Fig.~\ref{fig:RM_2D}) with well-defined PDFs for their model parameters (Fig.~\ref{fig:RM_PDF_indiv}), except for the first and last exposures. This is a consequence of the darkened flux of the stellar limb and its partial occultation by the planet, which results in a lower S/N at ingress and egress. For this reason, we exclude these exposures from further analyses. We note that the derived contrasts and FWHMs are consistent between the different retained exposures (Fig.~\ref{fig:RM_PDF_indiv}), as is expected from the stability of the G-type star. Surface RVs roughly go from $-$2 to $+$2 kms$^{-1}$ along the transit chord, meaning that the planet is successively blocking just as much blueshifted as redshifted stellar surface regions, which already suggests an aligned orbit and a moderate stellar rotation. 

\begin{figure}[h!]
    \centering
    \includegraphics[width=0.98\hsize]{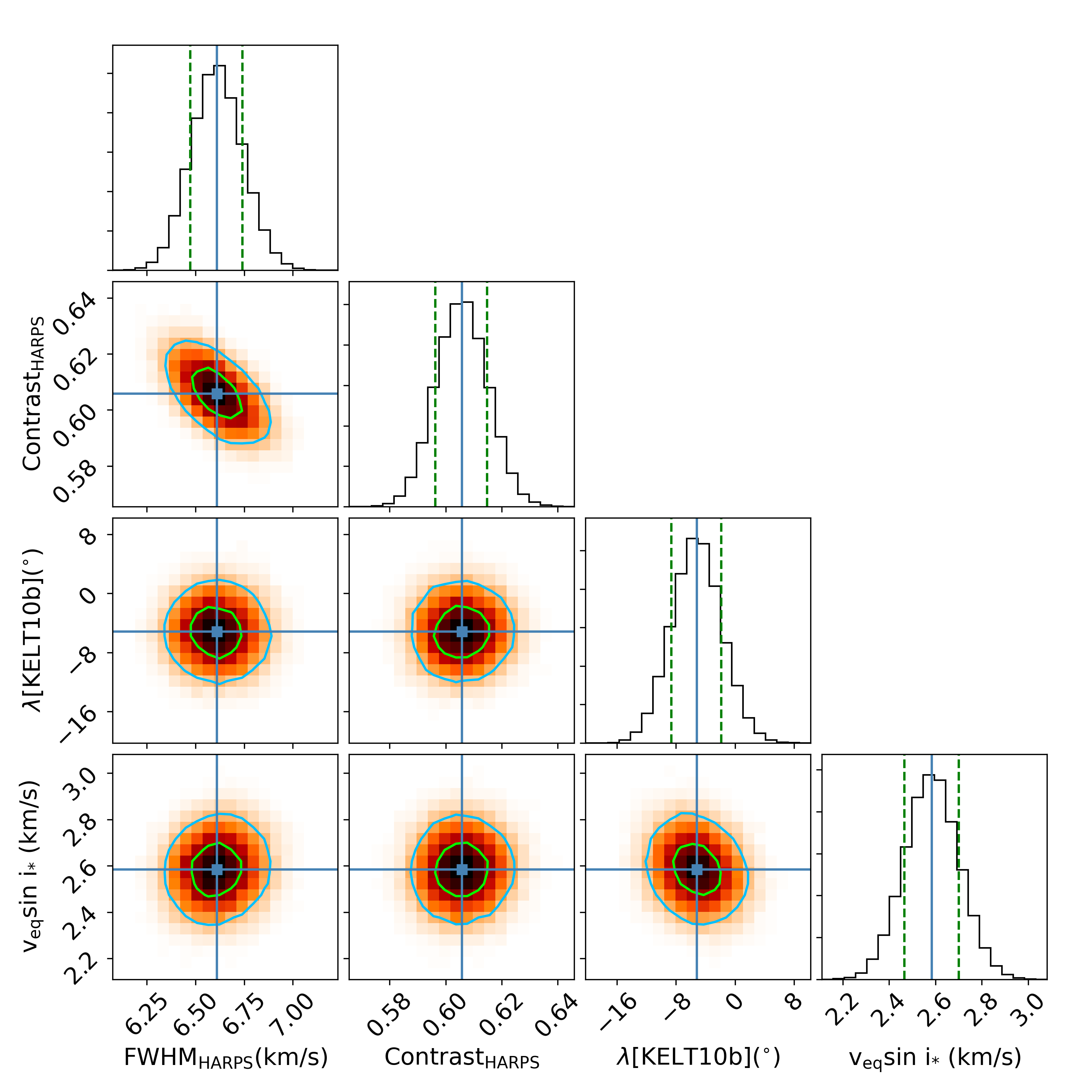}
    \caption{Correlation diagrams for the PDFs of the RMR model parameters for the KELT-10 b transit. Curved green and light blue lines show the 1 and 2$\sigma$ simultaneous 2D confidence regions that contain 39.3\% and 86.5\% of the accepted steps, respectively. One-dimensional histograms correspond to the distributions projected on the space of each line parameter, with the green dashed lines limiting 68.3\% of HDIs. The blue straight lines and squares show the median values.
    }
    \label{fig:RM_corner}
\end{figure}

\subsection{Derivation of the RM parameters}

\begin{table}
  \centering
\begin{tabular}{cccc}
\hline
Parameter & Unit & Reloaded & Revolutions (adopted)  \\
\hline
$\lambda$ & deg & $-6.4 \pm 3.7$ & $-5.2 \pm 3.4$ \\
$v_{\rm eq} \sin{i_\star}$ & kms$^{-1}$ & $2.61 \pm 0.13$ & $2.58 \pm 0.12$ \\
\hline
\end{tabular}
\caption{Derived parameters from the RM analysis, using two different methods.}
\label{tab:RM_results}
\end{table}

The RMR technique exploits all of the information contained in the transit time series by simultaneously fitting a model of the stellar line to all CCF$_{\rm intr}$. The local stellar line is modeled as a Gaussian profile with a constant FWHM and contrast, consistently with the analysis of the individual exposures (Fig.~\ref{fig:RM_PDF_indiv}). The time series of theoretical stellar lines was convolved with a Gaussian profile of width equivalent to the HARPS resolving power before being fitted to the CCF$_{\rm intr}$. Stellar solid-body rotation is assumed. The MCMC jump parameters are hence the line contrast and FWHM, the projected obliquity $\lambda$, and the projected equatorial stellar rotational velocity $v_{\rm eq} \sin{i_\star}$, with the same priors for the contrast and FWHM as in Sect.~\ref{sec_RM_indiv}, $\mathcal{U}(-180,180)^\circ$ for $\lambda$ and $\mathcal{U}(0,5)$ kms$^{-1}$ for $v_{\rm eq} \sin{i_\star}$.

The resulting best-fit line satisfyingly reproduces the local stellar lines along the transit chord, as can be seen in Fig.~\ref{fig:RM_2D}. The corner plot of the fit is shown in Fig.~\ref{fig:RM_corner}. The resulting PDFs are well defined and symmetrical, in a way that HDIs are here equivalent to quantile-based credible intervals. The correlation diagrams show no correlation between the various parameters, such that the system's architecture does not depend on the line model. Consistently with the individual exposure analysis (Sect.~\ref{sec_RM_indiv}), we find a low obliquity and stellar rotational velocity ($\lambda = -5.2 \pm 3.4^\circ$, $v_{\rm eq} \sin{i_\star} = 2.58 \pm 0.12$ kms$^{-1}$), and confirm that a constant stellar line model fits the data well. 

Given the quality of the visit, we further tried to constrain finer stellar effects. We fit for stellar differential rotation (DR) employing a law derived from the Sun $\Omega(\theta) = \Omega_{\rm eq} (1 - \alpha \sin^2\theta)$ \citep[e.g.,][]{cegla2016}, where $\Omega_{\rm eq}$ is the equatorial rotation rate, $\theta$ the latitude, and $\alpha \in [0,1]$ the unitless DR parameter. Compared to the solid-body fit, two more free parameters are considered here: $\alpha$ and $\sin{i_\star}$, for which we use noninformative priors. The resulting PDFs for $\alpha$ and $\sin{i_\star}$ are uniform within their definition intervals and the other parameters remain unchanged within 1$\sigma$, meaning that DR is not constrained. This is also supported by BIC comparison, which does not justify using the DR model over the solid body one.

For completeness, and to assess the stability of our results against the employed method, we also conducted a Reloaded Rossiter--McLaughlin \citep[RRM,][]{cegla2016} fit of the surface RVs. The RRM technique is restricted to the interpretation of the RV centroids from the planet-occulted CCFs. It requires that the surface RVs have a strong enough S/N to be fitted with a stellar line model, which is the case here. We assume solid-body rotation, and the same priors as for the RMR fit were used. The resulting parameters are consistent within 1$\sigma$ with those derived from the RMR analysis, indicating the robustness of our results. The various derived parameters can be found in Table \ref{tab:RM_results}.

\section{Transmission spectroscopy}
\label{sec_4}
This section describes the transmission spectroscopy performed on the HARPS dataset, including all the corrections done before the calculation. We followed the methods by \cite{mounzer2022}, which further refine methods by \cite{seager2000a}, \cite{redfield2008} and \cite{wyttenbach2015}.

We used the e2ds format of HARPS spectra (DRS version 3.5), which includes a 2D grid of 72 and 71 echelle orders (for fiber A and B, respectively), and each order has 4098 pixels. The first step in the analysis is deblazing, using the blaze calibration obtained for each night of observation. This study's primary focus is orders 56 and 67 (using the order numbering by DRS, 55 and 66 for the fiber B data). The difference in order numbering is due to a gap in fiber B on the order of number 45 (at 5277 \r{A})\footnote{Table 6.1 in: \url{https://www.eso.org/sci/facilities/lasilla/instruments/harps/doc/DRS.pdf}}. These orders include \Na\,and \Ha. 

We also excluded several additional spectra from the analysis observed at airmass>2. This filtering affected only the out-of-transit baseline and did not hamper our resulting master-out spectrum (the average out-of-transit spectrum used for stellar line correction). This filtering led to a final set of 37 spectra used in the analysis of transmission spectroscopy. 

\subsection{Correction for Earth's atmosphere}
\label{sub_sec_earth}
After correcting for the instrumental effects, we proceeded to mitigate the impact of  Earth's atmosphere on the observed spectra. To do so, we used several corrections. First, we used \texttt{molecfit} \citep{smette2015,kausch2015} to correct for the H$_2$O and O$_2$ lines from the atmosphere, following the procedures described in \cite{allart2017} and \cite{seidel2020a}. In a few words, \texttt{molecfit} takes an input scientific spectrum and fits it with a synthetic high-resolution (R$\sim$ 5 000 000) transmission spectrum in selected wavelength regions with prominent telluric absorption. It runs a line-by-line radiative transfer code using an extensive line list of atmospheric lines and calculates the transmission spectrum during the given observation. Finally, it corrects the scientific spectrum, significantly mitigating the effect of the telluric lines. A more detailed description of the program is in the references above.

After correcting for the H$_2$O and O$_2$ lines, we normalized the spectra using the median flux of each spectrum. This normalization still leaves a color effect in the spectrum due to the instrument that could be corrected by a better normalization approach or by subtracting a linear fit. Since this effect is stable for a night, it is unnecessary to rectify this trend since we will eventually divide by the master-out spectrum. 

After normalizing the spectra, we corrected for the cosmic rays. We calculated a median spectrum and flag pixels as cosmic (and set them to the median spectrum value) if the flux difference from the median spectrum was beyond the 5$\sigma$ level. We also manually checked the spectra for missed cosmics during the correction, finding that no cosmic rays hampered the region of \Na\,and \Ha.

Finally, \texttt{molecfit} does not correct for sodium emission in Earth's atmosphere. We monitored the fiber B data (aimed at the sky background) and found significant sodium emission. Our attempts to subtract this emission were unsuccessful because of the fiber A/B efficiency ratio variability at night. Due to this variation, we observed the remainder of the emission feature at high airmass, creating a spurious signal in the transmission spectrum. As a result, we thus opted to mask a small part (around 8 pixels) of the spectrum. 

Afterward, we shifted all the spectra into the rest frame of the star, using the BERV (approximately $2.7$ kms$^{-1}$ for night 1) and systemic velocity value calculated from the analysis of the master-out CCF (\Cref{sec_RM}; $31.96\pm 0.02$ kms$^{-1}$). The value of systemic velocity calculated from HARPS is compatible with both the values from \cite{kuhn2016} ($31.9\pm 0.1$ kms$^{-1}$) and the one used in \cite{mccloat2021} ($31.61\pm 1.29$ kms$^{-1}$).

\subsection{Correction for stellar effects}
\label{sec:stellareffects}
Once we shifted to the stellar rest frame, we could correct for the stellar lines. At first order, we could calculate the average out-of-transit spectrum ($M_{out}$) and divide our dataset by it \citep{wyttenbach2015,seidel2019}. However, doing so disregards the RM and center-to-limb variation (CLV) effects that can affect our spectra. This approach is the one \cite{mccloat2021} took, as the lack of precise wavelength calibration in their UVES data disallowed the measurement of RVs from the spectra, preventing an RM correction. HARPS, however, is designated to provide accurate RVs, hence we could measure and correct for the RM effect and simulate its impact on the final transmission spectrum.

To correct our spectra for the RM effect, we adapted the formula by \cite{mounzer2022}: 
$$
F_\mathrm{corr,i} = 1 + \left(\frac{R_p^2}{R_s^2}\right) \left(1 - \frac{M_\mathrm{out} - F_i \delta_i}{M_\mathrm{out,v_i}\Delta_i}\right).
$$
Here, $\frac{R_p^2}{R_s^2}$ corresponds to the planet/star radius ratio, $F_i$ is the spectrum we wish to correct,  $M_\mathrm{out}$ is the master-out spectrum, and $M_\mathrm{out,v_i}$ is master-out shifted by the local stellar velocity. The $\delta_i$ and $\Delta_i$ correspond to the transit depth and white-light flux received during the individual exposure. The $v_i$ velocity was calculated using the \obliquity and \stelprojvel\, values calculated in \Cref{sec_RM}, using an analytical formula (for circular orbit) describing the RVs of the stellar surface along the transit chord of the planet for each exposure:
\begin{equation}
    \resizebox{.48\textwidth}{!}{$v_{i}(\phi) = \left(\frac{a}{R_s}\right) (\sin(2 \pi \phi) \cos(\lambda) + \cos(2 \pi \phi) \cos(i) \sin(\lambda)) v_{\rm eq} \sin{i_\star,}$}
    \nonumber
\end{equation}

where $\phi$ is the orbital phase at which we observed the planet. Other parameters follow the same notation as in previous sections (cf. \autoref{Tab_conan_output} and \Cref{sec_RM}). The shift is between $-2.6$ to $+2.6$ kms$^{-1}$. This roughly corresponds to 3-4 pixels (1 pixel $\sim 820$ms$^{-1}$) in HARPS spectra from the unshifted $M_\mathrm{out}$.

To estimate the impact of the RM effect on the transmission spectrum of KELT-10 b (for comparison with the \cite{mccloat2021} detection), we used the EVaporating Exoplanets (\texttt{EVE}) code \citep[e.g.,][]{bourrier2013,allart2018}. This code allows for simulations of the 3D atmosphere of a transiting planet and calculations of synthetic transmission spectra accounting for the geometry of the transit and the properties of the stellar surface. First, we accounted for CLV of the local stellar spectrum by tiling the EVE stellar grid using synthetic line profiles calculated at different limb darkening angles with the \texttt{Turbospectrum}\footnote{\url{https://github.com/bertrandplez/Turbospectrum2019}} code \citep{plez2012}. Then the RM effect was naturally accounted for by shifting the local stellar spectra according to the rotational velocity of KELT-10. Using these settings, we simulated a transit of KELT-10b without including any atmosphere to assess the distortion induced purely by the RM effect and CLV. We set the code to return a time series of KELT-10 spectra as they would be observed with HARPS and we used them to compute an average transmission spectrum comparable to our observations.

\begin{figure*}
    \centering
    \includegraphics[width=0.95\hsize]{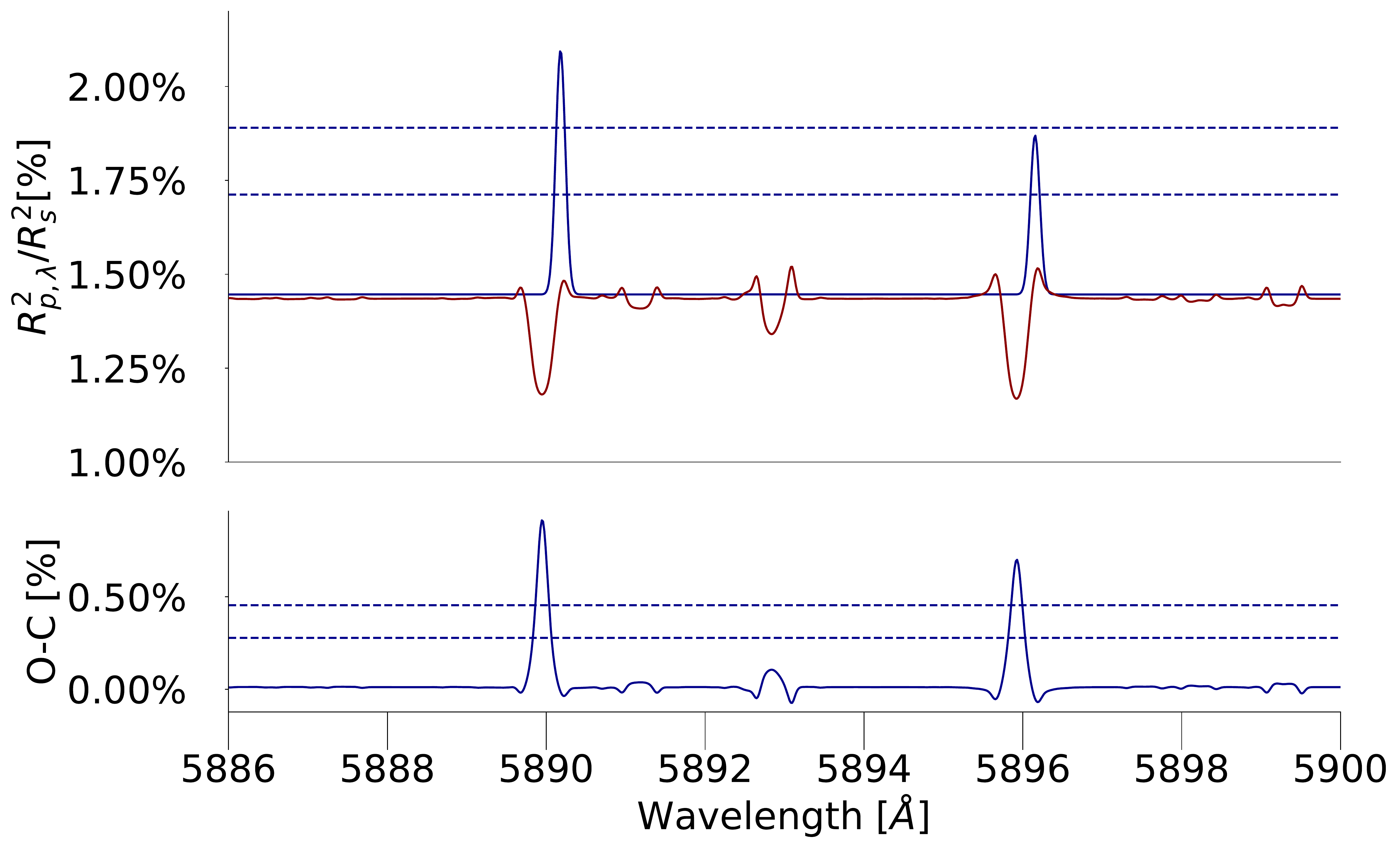}
    \caption{Modeling of the RM + CLV effect on the transmission spectrum in the rest frame of the planet, specifically focused on the sodium detection presented in \cite{mccloat2021} ($A_{D2}=0.66\%$, $\sigma_{D2}=0.06\r{A}$, $\Delta\lambda_{D2}=+0.23\r{A}$; $A_{D1}=0.43\%$, and $\sigma_{D1}=0.06\r{A}$, $\Delta\lambda_{D1}=+0.23\r{A}$). Top panel: Gaussian fit to the lines attributed by \cite{mccloat2021} to planetary sodium (their Fig. 4, blue) and simulated RM and CLV effects using EVE (red). We note that compared to \cite{mccloat2021}, we use a reversed y-axis (due to different units used), that is absorption is above continuum and emission is below it. The $3\sigma$ and $5\sigma$ level reported by \cite{mccloat2021} is shown as dark blue dashed lines. Bottom panel: O-C values (UVES - EVE simulation, blue), and $3\sigma$ and $5\sigma$ UVES detection levels (blue).
    }
    \label{Fig_RM_effect_ts}
\end{figure*}

\autoref{Fig_RM_effect_ts} shows the comparison between our EVE simulation and the best-fit model to the UVES transmission spectrum from \cite{mccloat2021} in the region of the sodium doublet. We are unable to explain the observed signal with the pure RM and CLV effects. The amplitude of the effect is on the order of $0.2\%$, and opposite to an atmospheric absorption signal, impacting the transmission spectrum significantly. It is even possible that the RM effect is obscuring part of the putative atmospheric signal; however, given the issues with the wavelength calibration in the UVES data \citep[Section 5.3 in ][]{mccloat2021}, it is difficult to draw a definite conclusion.

If the observed UVES signal is truly redshifted, we could explain the "emission-like" shoulders apparent in the UVES transmission spectrum \cite[Fig. 4 in ][]{mccloat2021}. However, only the D1 signal there seems to be as broad as our simulation, while the D2 signal appears to be due to a few pixels. The amplitude of the signature would be slightly reduced, but not enough to compromise its detection.

On the other hand, if the redshift is purely of nonastrophysical origin, the RM effect would significantly decrease the observed signal, meaning the true amplitude of the sodium signal is higher. In that case, KELT-10 b would have one of the strongest sodium signals detected yet, which, considering its location in the hot Jupiters' population (\autoref{Fig_radius_insolation}), seems curious, as we would expect a stronger signal in hotter and puffier planets (as long as the sodium itself is not ionized). It is also possible that due to the lower resolution of UVES spectra, the RM effect is smeared out and hence weaker.

After stellar line correction, including the RM effect (CLV is negligible compared to RM), a visible \Na\,remnant is still present. This remnant results from very low S/N in the core of the stellar \Na\,and should be removed, as it can create spurious signals. We masked the core region with NaN values, following the procedures presented in \cite{seidel2020,seidel2020a}. This way, we could mask the sodium remnant, but we unfortunately also masked part of the planetary signal. 

To estimate how much of the planetary signal was removed, we used the value of the mean FWHM of the \Na\,signal from \cite{seidel2019} (FWHM = 0.65$\r{A}$). The area within the FWHM marked the region for the planetary signal, and we assumed no shift of the sodium line (in the planetary rest frame). The fraction of information lost is $\frac{FWHM_\mathrm{mask}}{FWHM_\mathrm{signal}}=28\%$. After we shifted the spectra to the rest frame of the planet, we calculated an average transmission spectrum over full in-transit exposures, weighting them by S/N.
\begin{figure*}[]
    \centering
    \includegraphics[width=0.8\hsize]{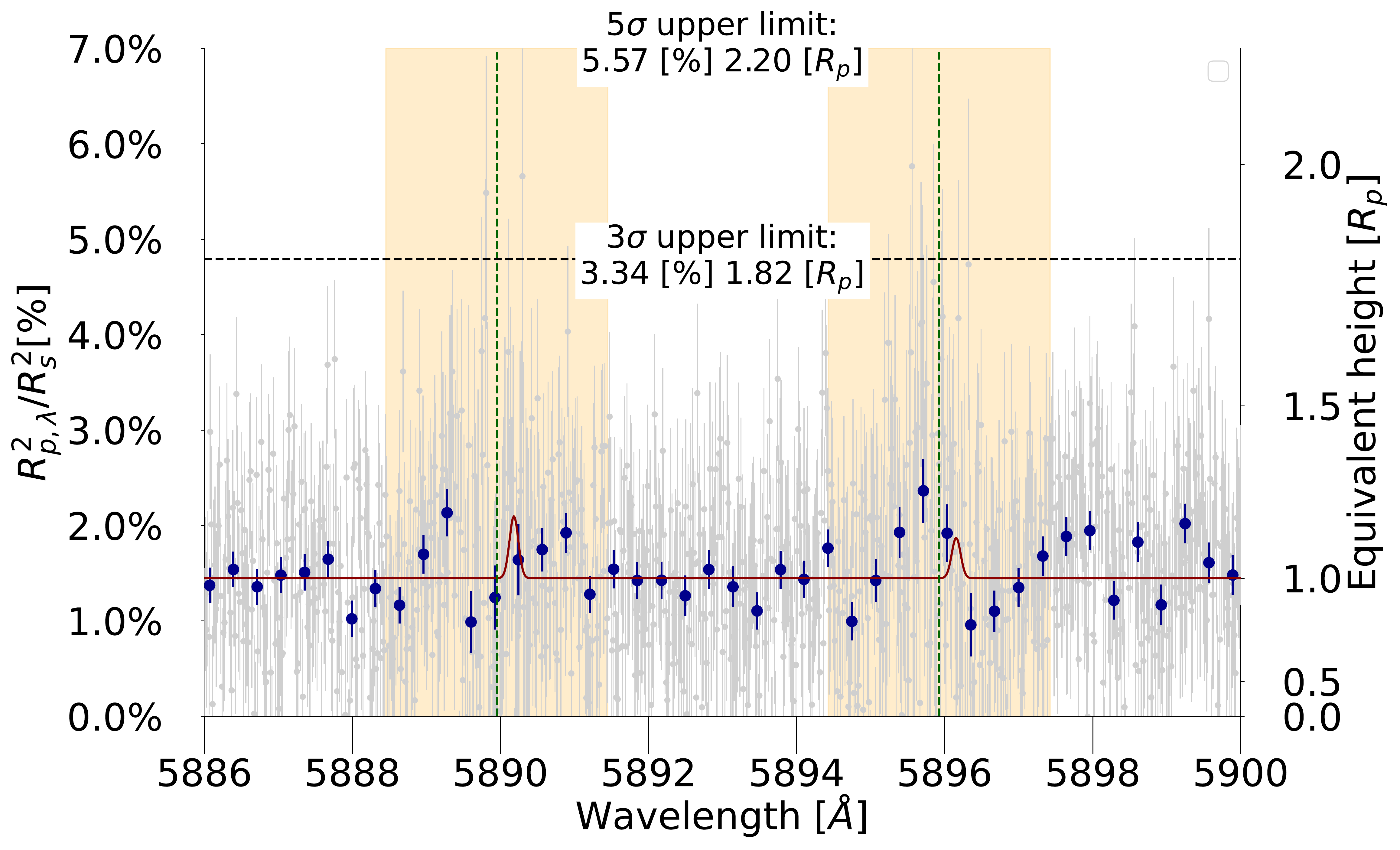}
    \caption{Transmission spectroscopy of KELT-10 b in the region of the sodium doublet. In gray, the entire transmission spectrum is shown (corrected for RM effect, as described in \autoref{sec:stellareffects}), with 20x bins in dark blue. The detection of \cite{mccloat2021} is overplotted (dark red) as a visual cue for comparison. The black dashed line corresponds to the 3$\sigma$ and 5$\sigma$ limit above the continuum ($R_{p}^2/{R_s^2} \sim 1.44 \%$). The yellow area is the $\pm$1.5 $\r{A}$ region around each of the \Na\,lines.
    }
    \label{Fig_trans_na_abs}
\end{figure*}

\begin{figure*}[!h]
    \centering
    \includegraphics[width=0.8\hsize]{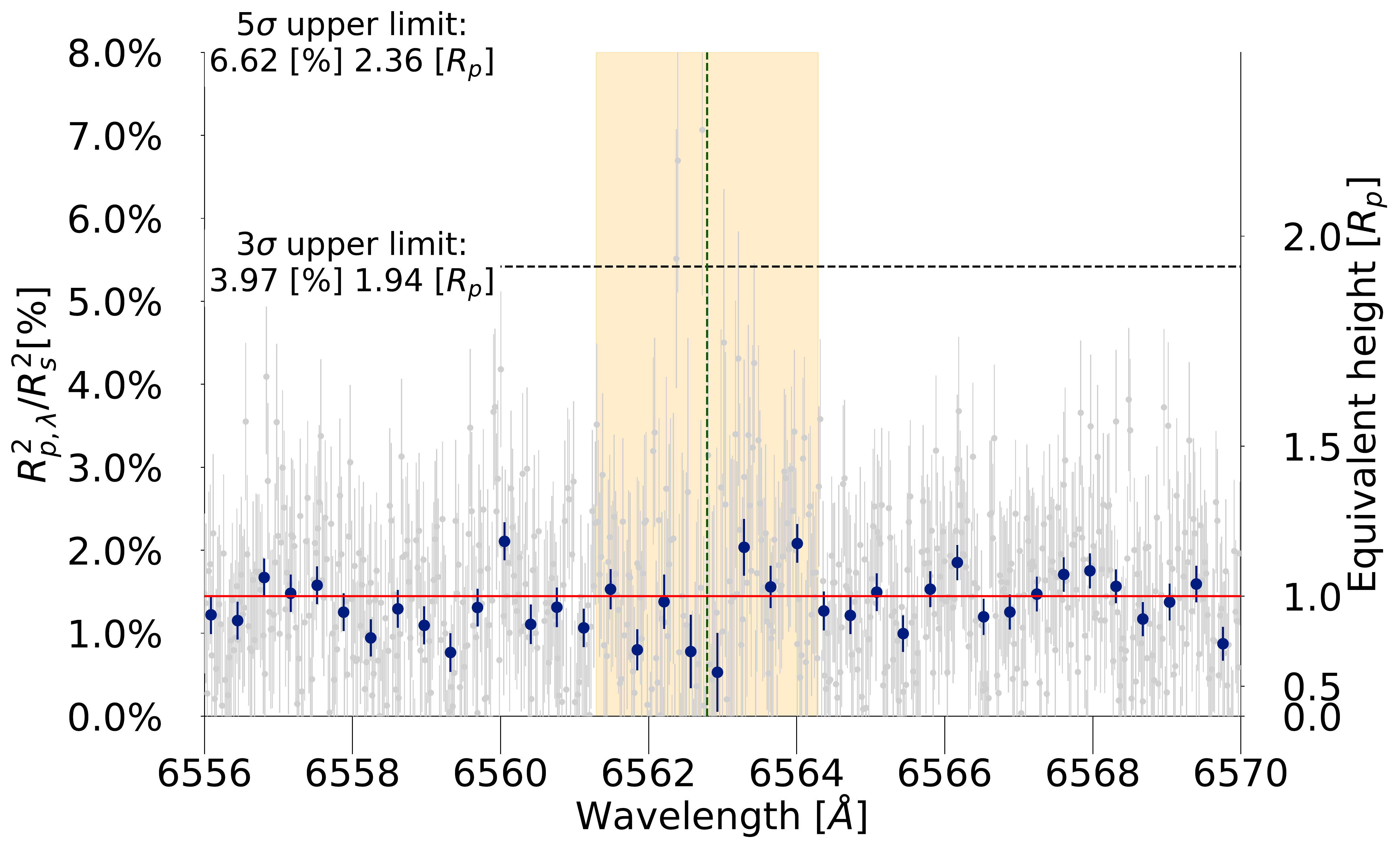}
    \caption{Same as \autoref{Fig_trans_na_abs}, but in the region of \Ha. The color-coding is the same as in Fig. 8. The additional red line is a continuum of one equivalent height. No detection is visible in this region either, with corresponding 3$\sigma$ and 5$\sigma$ limits on the plot. 
    }
    \label{Fig_trans_ha_abs}
\end{figure*}

\subsection{Transmission spectroscopy}
The final transmission spectrum is shown in \autoref{Fig_trans_na_abs} and \autoref{Fig_trans_ha_abs} for \Na\,and \Ha, respectively. Although we corrected our spectrum for the RM effect, this effect is negligible within our detection limits. We also checked the RM uncorrected spectrum, finding no significant changes in the notable features of the spectrum.

Neither of the sodium lines were detected. As part of the comparison with \cite{mccloat2021}, a Gaussian model with their reported parameters has been over-plotted. To calculate the upper limit, we took the mean uncertainty in the $\pm$1.5 $\r{A}$\,region around the lines, marked as a yellow area, and used it as a $1\sigma$ estimate. We used a broader region to estimate the uncertainty since we do not know of any potential shift of the possible signal. 
These kinds of shifts can occur due to wind patterns and this is quite common \citep[e.g.,][]{snellen2010,seidel2020b,seidel2021,mounzer2022}. The $3 \sigma$ upper limits are \Naupperlimvaluewounit\, for \Na\,and \Haupperlimvaluewounit\,for \Ha. These upper limits are likely underestimated due to the distortion of the putative atmospheric signal by the planet-occulted stellar lines.

In the region of \Na,\,we see a few bins that show absorption. However, the location of these bins in each of the doublet lines is different, with -30kms$^{-1}$ for the D2 feature and -10kms$^{-1}$ for the D1 feature. This suggests that these features are of nonastrophysical origin. In fact, the D2 feature can be attributed to telluric sodium. For the D1 feature, we attribute this to a low S/N remnant (the count in the line core is below 100) and systematics in the measurements -- several features of a similar strength have been observed (in the red part of the spectrum) as well. These features also hint at potential correlated noise. Finally, this kind of signal, if real, would be visible in the UVES data. Based on these arguments, we, therefore, claim nondetection.

\section{Conclusion}
\label{sec_5}

\subsection{Bulk planetary and orbital properties}
We have carried out an independent photometry analysis, which shows consistent results with parameters from \cite{kuhn2016} within $1\sigma$. We improved the precision of the P and $T_{14}$ parameters by a factor of $\sim 18$.

We were able to study the RM effect with great precision using the RMR method \citep[][]{bourrier2021}, finding the orbit of KELT-10 b to be aligned with the stellar rotation plane, in projection onto the plane of the sky \obliquity. The system is likely truly aligned, as previous results show that hot Jupiters around late-type stars ($T_{\rm eff} < 6250$ K) tend to live on aligned orbits \citep{winn2010,albrecht2022}. Indeed, late-type stars are expected to have substantial convective envelopes, which trigger stronger tidal interactions, thus realigning their systems more efficiently. This argument is supported by the proximity of KELT-10 b to its host star ($a/R_\star < 10$, \autoref{Tab_conan_output}), and by the relatively old age of the system \citep[$4.5 \pm 0.7$ Gyr,][]{kuhn2016}.

\subsection{Transmission spectroscopy}
The nondetection of \Naupperlimplanrad\,(\Na) and \Haupperlimplanrad\,(\Ha) corresponds to no extreme extension of upper atmospheric layers, as expected from such a "well-behaved" system. The RM + CLV effects create a spurious signal, very similar to the one detected by \cite{casasayas-barris2021} on the example of HD 209458 b. In future studies of this system, these effects should be corrected for in the transmission spectra or included in EVE-like simulations to derive (a) confident atmospheric detection(s). 

\subsection{Comparison with previous sodium detection}
Recently, \cite{mccloat2021} have claimed sodium detection from a VLT/UVES single transit dataset. However, these data could not allow these authors to obtain RV measurements precise enough to correct for the RM effect (which was attributed to \texttt{molecfit} being unable to fine-tune the wavelength grid --  see section 4.6 in \cite{mccloat2021}). 

Since the authors could not measure the RVs during the transit, a potential RM effect could have created a significant signal, as in the case of \cite{casasayas-barris2021}. Furthermore, the tentative sodium detection had a significant redshift, likely due to the wavelength calibration issues mentioned above.

We modeled the effect of RM and CLV on the transmission spectrum, finding that the distortion by planet-occulted stellar lines cannot explain the measured sodium detection. This discrepancy with the detection of \cite{mccloat2021} is surprising, especially if the observed redshift is instrumental, as the RM effect would significantly strengthen the sodium detection, making it one of the strongest detections of the \Na\,doublet to date. 

Our transmission spectrum cannot confirm the UVES sodium detection \citep{mccloat2021}, but we also cannot rule it out, as the lower quality of the final transmission spectrum does not allow us to probe to such detection limits. There are several potential reasons why our results do not fully match. 

First, our RM correction method might introduce several biases. In this work, we did not thoroughly investigate different stellar models to use in the RM+CLV modeling. Incorrect RM correction can then introduce spurious effects that blur the signal (under-correction) or create a fake one (over-correction). 

Furthermore, \texttt{molecfit} struggles with the correction of low S/N spectra, which is the case locally in the core region of the sodium doublet. In addition to that, due to low counts on the detector in the stellar core (<100), we are no longer photon-noise dominated but red-noise dominated (\cite{seidel2022,pepe2021}, see \autoref{sec_4} and \autoref{sec:stellarmask}). This is the case for all observations when the observed planetary signal is on the order of (or less than) a single photon on the detector and needs to be treated carefully. 

Finally, it is possible that either this work or \cite{mccloat2021} is further affected by some unaccounted effect (e.g., stellar variability, instrument issue), which could cause a discrepancy between the observations. In fact, the peaks in our transmission spectrum further in the red hint at the possibility of correlated noise and systematic errors, which are also on the same order as the expected signal based on UVES \citep{mccloat2021}. On the other hand, atmospheric variability can significantly affect the dataset, especially for slit-based spectrographs such as UVES.


We conclude that to properly address the question on the presence of chemical species in the atmosphere of KELT-10b (and confirmation of the previous \Na\,detection), we need additional data from larger telescopes than the 3.6 m telescope at La Silla to obtain a good S/N and temporal sampling and to correct for the stellar effects such as RM that cause a significant impact ($\sim0.2\%$) on the transmission spectrum. 
 
\begin{acknowledgements}
We would like to thank the anonymous reviewer for the insightful comments. We would like to thank Sean McCloat, who was kind enough to provide reduced UVES data for comparison purposes with the analysis presented in this paper. This project has received funding from the European Research Council (ERC) under the European Union’s Horizon 2020 research and innovation programme (project {\sc Four Aces}; grant agreement No 724427).  It has also been carried out in the frame of the National Centre for Competence in Research PlanetS supported by the Swiss National Science Foundation (SNSF). DE and MS acknowledges financial support from the SNSF for project 200021\_200726. OA and VB's work has been carried out in the frame of the National Centre for Competence in Research PlanetS supported by the Swiss National Science Foundation (SNSF). OA and VB acknowledge funding from the European Research Council (ERC) under the European Union’s Horizon 2020 research and innovation programme (project {\sc Spice Dune}; grant agreement No 947634). This project has also been carried out in the frame of the National Centre for Competence in Research PlanetS supported by the SNSF. M.L. acknowledges support of the Swiss National Science Foundation under grant number PCEFP2\_194576. This project has received funding from the European Research Council (ERC) under the European Union’s Horizon 2020 research and innovation program (grant agreement No 742095 ; SPIDI : Star-Planets-Inner Disk-Interactions; \url{http://www.spidi-eu.org}). S.G.S acknowledges the support from FCT through the contract nr.CEECIND/00826/2018 and POPH/FSE (EC)
\\
Python packages used: \texttt{numpy}, \texttt{scipy}, \texttt{matplotlib}, \texttt{astropy}\footnote{\cite{astropycollaboration2022}}, \texttt{specutils}, \texttt{batman}

\end{acknowledgements}

%
\bibliographystyle{aa} 
\bibliography{biblio} 
%
\begin{appendix}

\section{Observational log}
\begin{figure}[t!]
    \centering
    \includegraphics[width=\hsize]{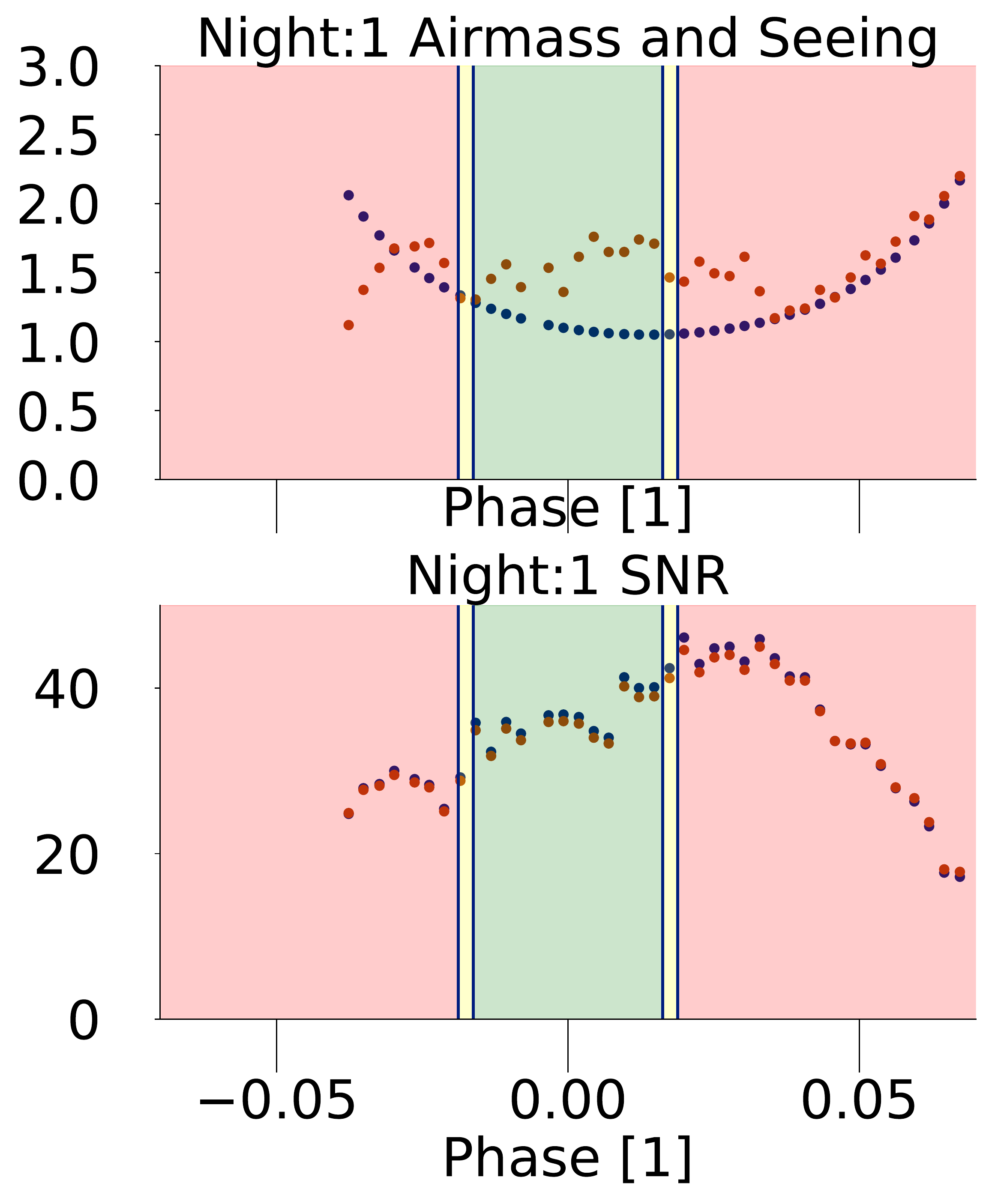}
    \caption{Observational log for the first night (26 June 2016). Top panel: Airmass (blue) and seeing (red). Bottom panel: S/N for order number 56 (blue) and 67 (red).}
    \label{Fig_night_1}
\end{figure}

\begin{figure}[t!]
    \centering
    \includegraphics[width=\hsize]{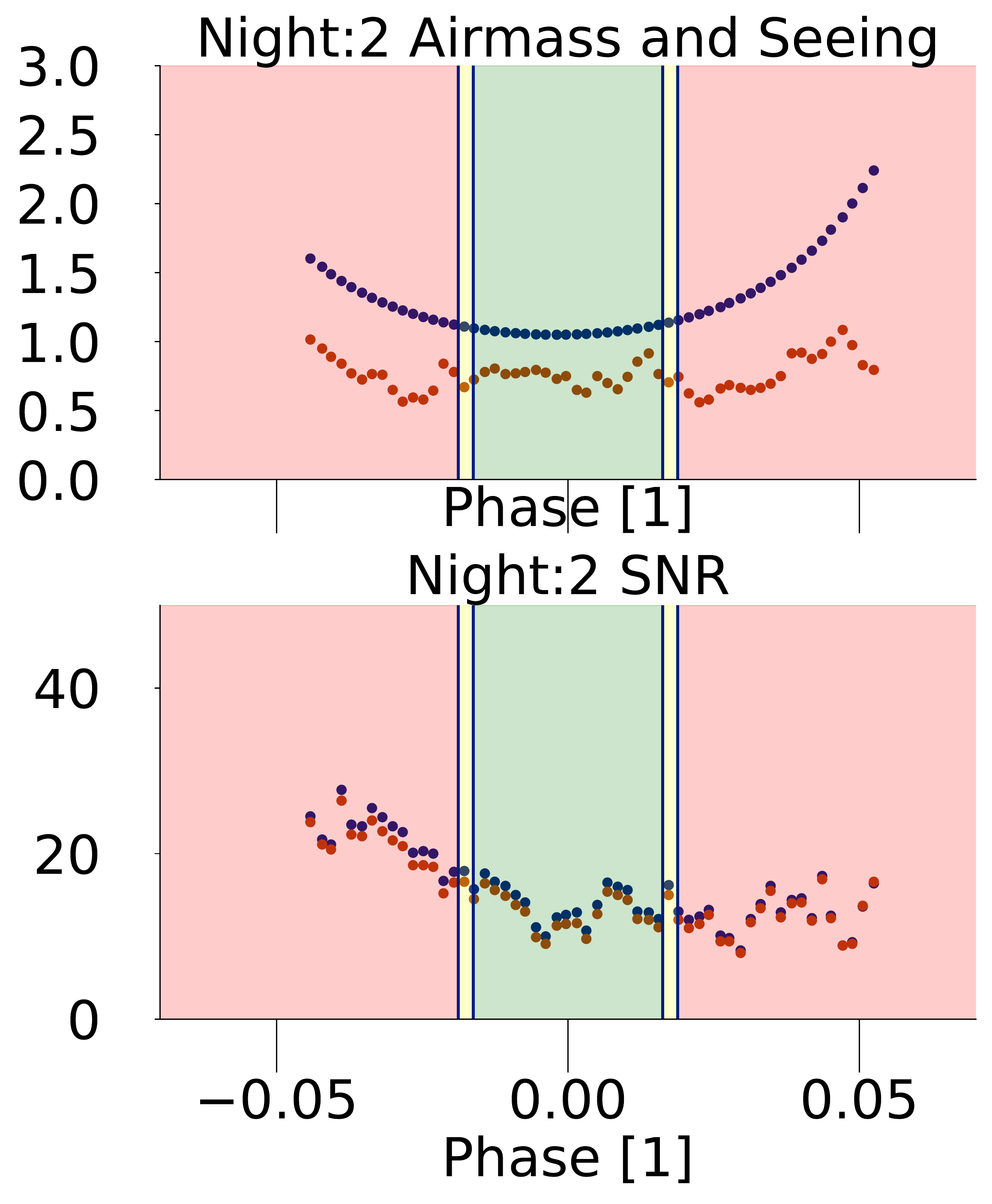}
    \caption{Same as \autoref{Fig_night_1}, but for the second night (21\ July 2016).
        }
    \label{Fig_night_2}
\end{figure}

Detailed observational log for both nights of HARPS. The first night (\autoref{Fig_night_1}) suffered from high seeing, which varied during the night. The second night (\autoref{Fig_night_2}) shows a low S/N, leading to removal from the dataset.

\section{Fiber B sodium emission}
\begin{figure}[!h]
    \centering
    \includegraphics[width=\hsize]{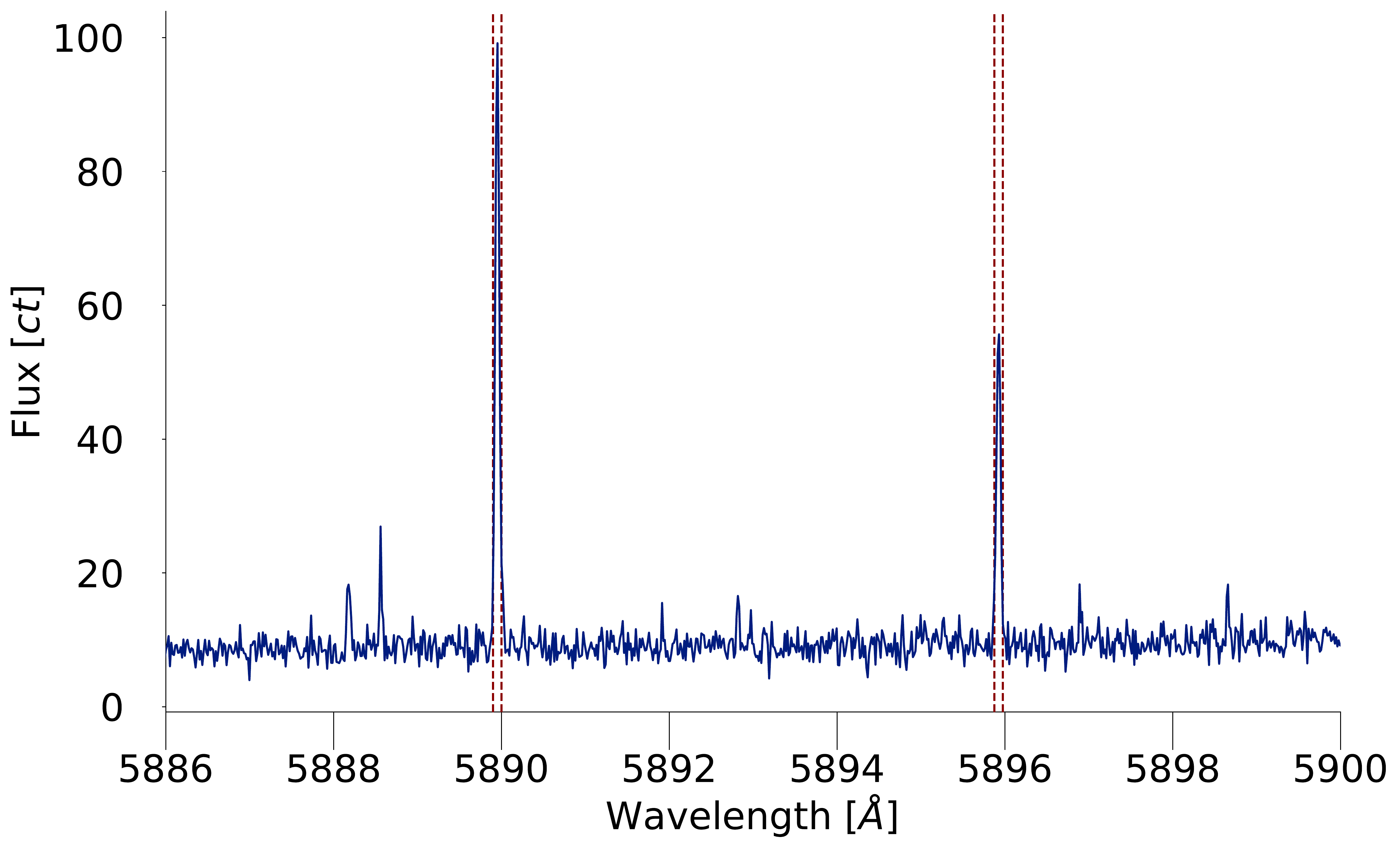}
    \caption{Sodium emission observed on fiber B during the first night of observation. Not correcting for it causes a significant absorption feature. No features were found on fiber B for the wavelength region of \Ha. The red dashed line signifies the masked part of the emission.
        }
    \label{Fig_fiber_B_sodium}
\end{figure}

As discussed in \Cref{sub_sec_earth}, we detected a significant telluric emission feature from sodium. This is shown in \autoref{Fig_fiber_B_sodium}, on a total master spectrum of the first night of fiber B. 
\clearpage

\section{RM analysis of the discarded visit}
\label{app:RM}

\begin{figure}
    \centering
    \includegraphics[width=\hsize]{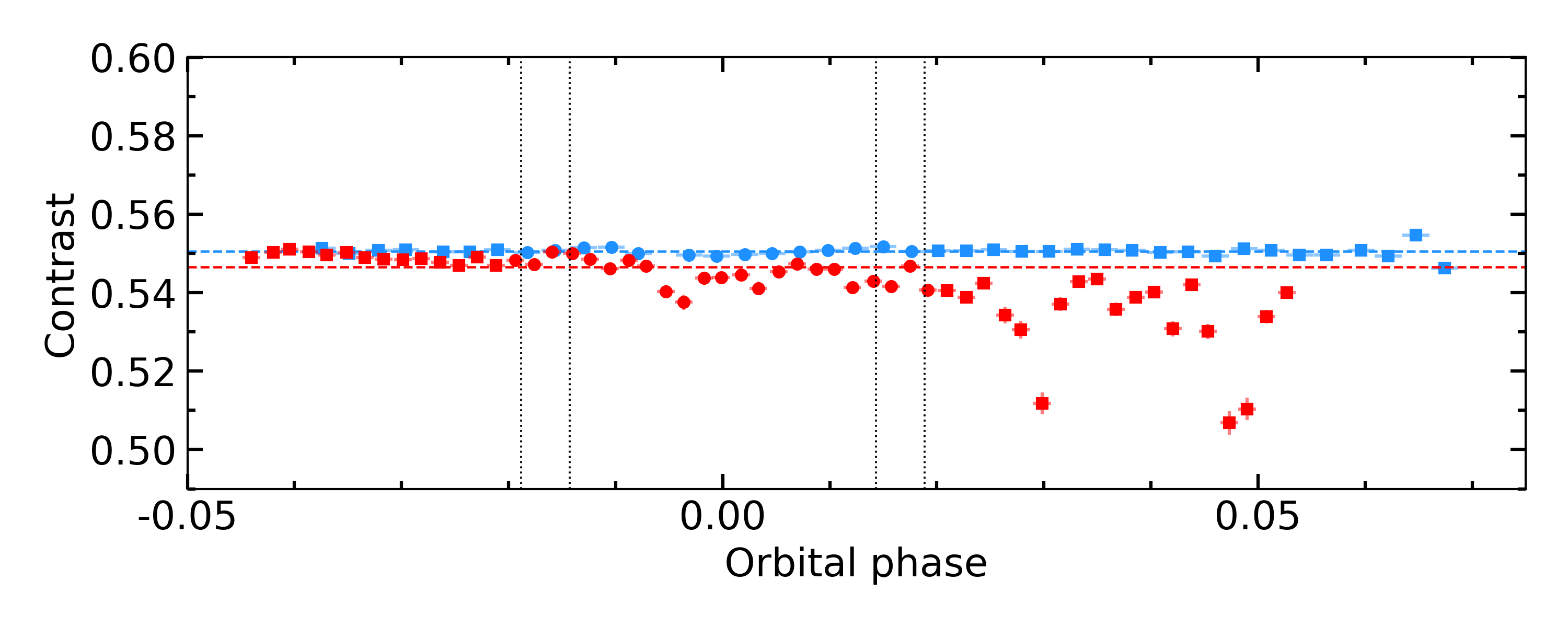}
    \includegraphics[width=\hsize]{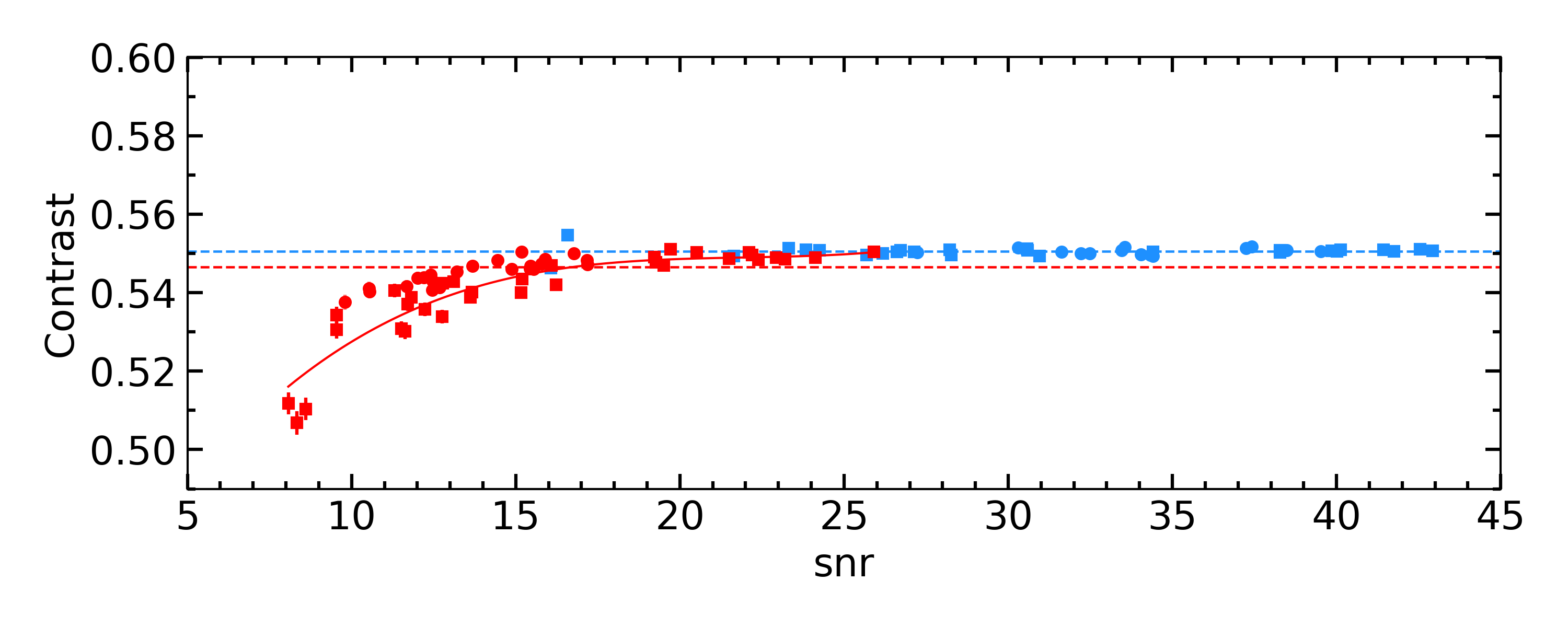}
    \caption{Contrast of the disk-integrated CCFs as a function of orbital phase (top) and S/N (bottom) for the 26 June 2016 visit (blue points) and the 21 July 2016 visit (red points). The horizontal dashed lines indicate the mean value for each visit. The vertical dashed lines represent the transit contacts. Out-of-transit (in-transit, resp.) exposures are shown as squares (circles, resp.). The full-line curve in the lower plot represents the best polynomial fit.
        }
    \label{fig:RM_raw}
\end{figure}

An additional transit was recorded with HARPS on 21 July 2016. The S/N of this visit is substantially lower than the 26 June 2016 one (16 on average for the former versus 31 for the latter, see Fig.~\autoref{Fig_night_1} and \autoref{Fig_night_2}). Furthermore, the 21 July 2016 spectra were Moon-contaminated shortly after the end of the transit, resulting in spurious features in the affected CCFs. This is visible in the disk-integrated contrast fluctuations starting after the transit (Fig.~\ref{fig:RM_raw}, upper panel). We show the contrasts of the 26 June 2016 CCFs for a visual comparison of the stability of the two nights. 

\begin{figure}
    \centering
    \includegraphics[width=\hsize]{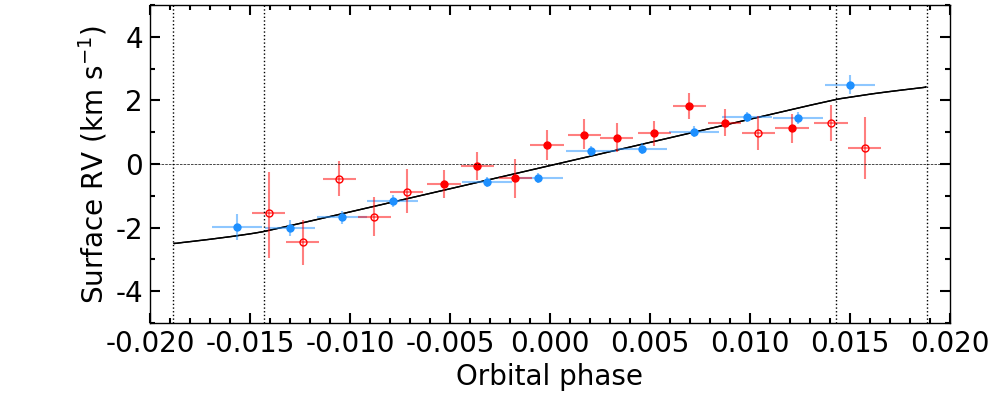}
    \includegraphics[width=\hsize]{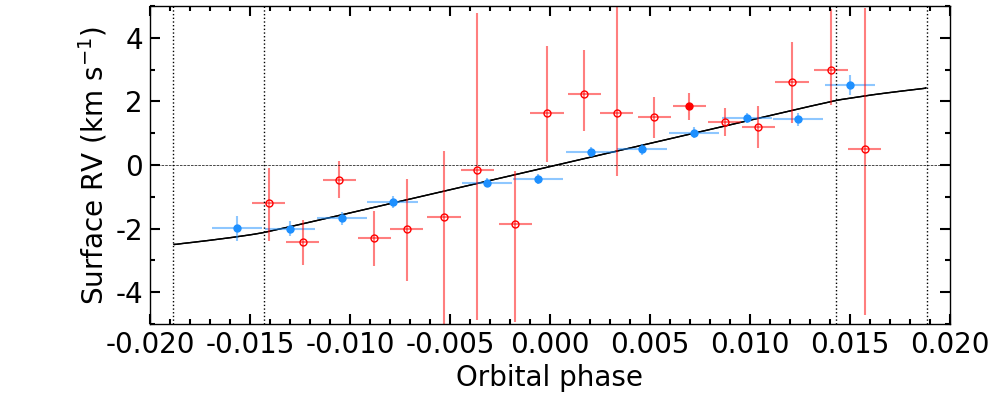}
    \caption{Intrinsic RVs as a function of orbital phase before (top) and after (bottom) the correction for the contrast--S/N correlation, for the 26 June 2016 visit (blue points) and the 21 July 2016 visit (red points). The horizontal dashed line indicates the zero RV level. The vertical dashed lines represent the transit contacts. The solid black line is the analytical RV curve for the adopted values of Table~\ref{tab:RM_results}.
        }
    \label{fig:RM_rv_ph}
\end{figure}

The 21 July 2016 fluctuations correlate with the S/N, as seen in Fig.~\ref{fig:RM_raw} (lower panel). We have captured this correlation by fitting a third order polynomial to the affected visit, based on BIC comparison. We then corrected the disk-integrated (DI) CCFs of this visit according to the fit result, which reduced the dispersion around the mean contrast value by a factor of $\sim$ 3. Ultimately, this correction turned out to be detrimental to the intrinsic CCFs. Indeed, Fig.~\ref{fig:RM_rv_ph} shows a large dispersion of the derived intrinsic RVs after correction, which results in very large error bars for the derived RM parameters. We have thus discarded the correction.

We further tried to circumvent the Moon contamination by excluding the post-transit baseline from the construction of the master out. Expectedly, the systemic velocity derived from the master out now coincides 26 times better between the two nights. Yet, the RMR analysis after this correction yields a very large difference in the derived parameters between the two nights. The same settings for the MCMC exploration as in Sect.~\ref{sec_RM} were used. The best-fit results for the affected visit are $v_{\rm eq} \sin{i_\star} = 1.88 \pm 0.21$ kms$^{-1}$ and $\lambda = 23.0^{+8.7}_{-9.0}{^\circ}$. Due to the 2.9-$\sigma$ discrepancy between the two nights for the obliquity value, we have also discarded the correction.

As a last attempt, we excluded the Moon-affected regions from the DI CCFs (roughly [-20 , 0] kms$^{-1}$), which should represent a smoother correction than excluding the full post-transit baseline. Still, a significant discrepancy in the RMR results between the two nights persists, even if it is lower than with the last correction: $v_{\rm eq} \sin{i_\star} = 2.43^{+0.21}_{-0.22}$ kms$^{-1}$ and $\lambda = 9.1^{+8.8}_{-9.1}{^\circ}$. Even if the stellar equatorial velocities are now compatible, the obliquities have a 1.5-$\sigma$ discrepancy between the two nights. In the absence of agreement between the two visits and despite our tentative corrections, we have discarded the Moon-contaminated night.
\section{Masking stellar sodium remnant}
\label{sec:stellarmask}
\begin{figure*}[!h]
    \centering
    \includegraphics[width=0.9\hsize]{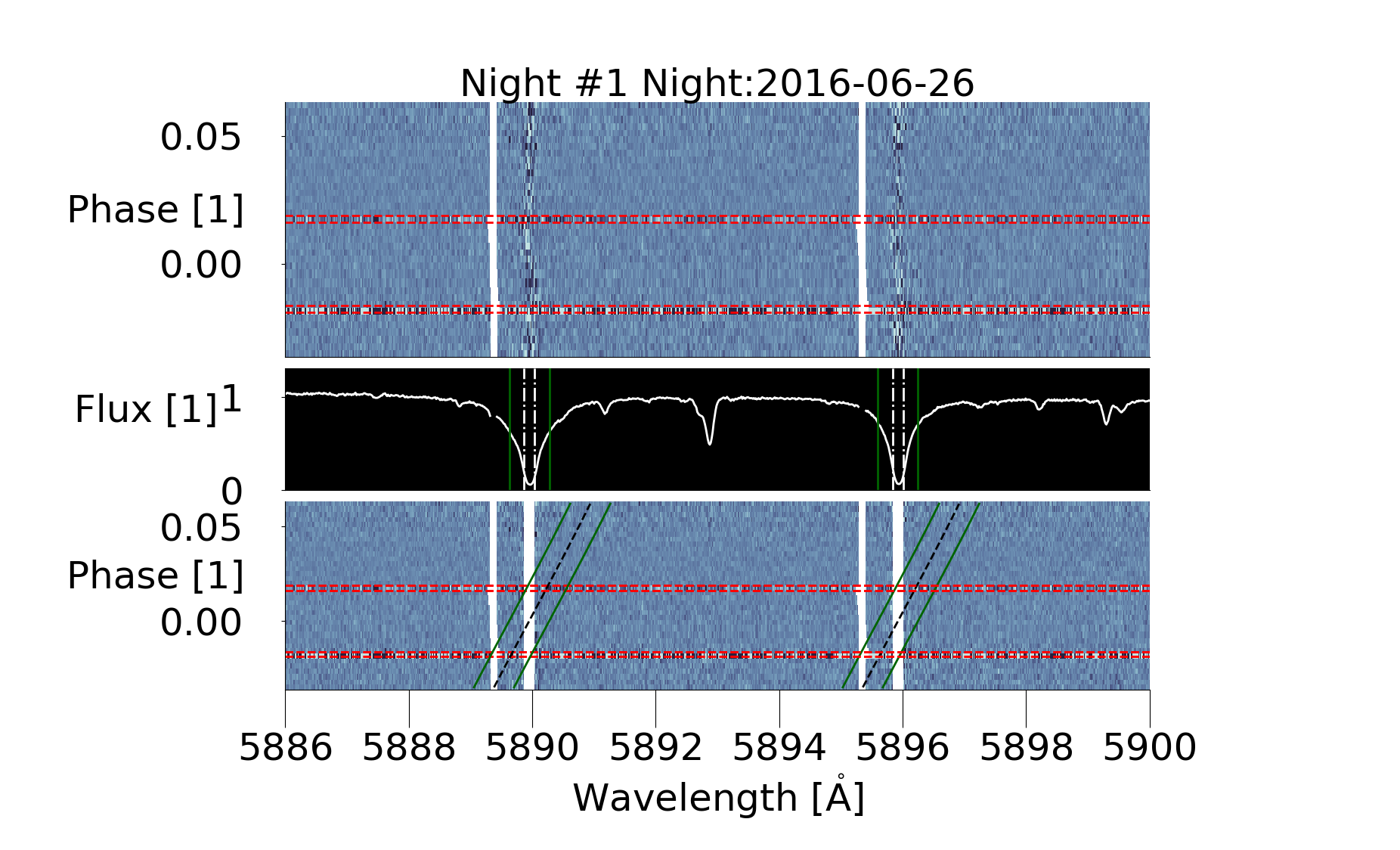}
    \caption{Used mask in the steps as described in \Cref{sec_4}. Top panel: Colormap of all spectra during the first night in the rest frame of the star. The red dashed lines correspond to transit contact points. The white mask is the mask used for telluric sodium, located at roughly $-30$kms$^{-1}$. Middle panel: Master-out spectrum, with the following colored lines: white dash-dot line, masked region; and green, sodium line $\pm \frac{FWHM}{2}$ region as reported in \cite{seidel2019}. Bottom panel: Masked stellar signal based on the mask shown in the middle panel (white lines), which is still in the rest frame of the star. The black line corresponds to the sodium line in the planetary rest frame, considering no additional shifts (e.g., due to winds). Green lines are offset by $\pm \frac{FWHM}{2}$ as used in the middle panel, and they correspond to the area from which we estimated our lost signal. The noisy spectra at ingress and egress (around red dashed line) are due to RM correction and light-curve weighting. These two spectra were not used in the final transmission spectrum.
        }
    \label{Fig_mask_na}
\end{figure*}

After correcting for master out, we observed a low S/N remnant in the sodium doublet due to line saturation. To remove this effect, we masked the region, as shown in \autoref{Fig_mask_na}

\end{appendix}

\end{document}